\documentclass[12pt, draftclsnofoot, onecolumn]{IEEEtran}

\usepackage{epstopdf}
\usepackage{multirow}
\usepackage{amsmath,epsfig,verbatim,amsopn,subfigure,color}
\usepackage{cite,xspace}
\usepackage{array}
\usepackage[ruled,lined,boxed,commentsnumbered]{algorithm2e}
\usepackage{array}
\usepackage{multirow,array}%
\usepackage{multicol}%
\usepackage{setspace}
\usepackage{here}
\usepackage{amssymb}
\usepackage{dsfont}
\usepackage{subfigure}
\usepackage{soul}

\usepackage{breqn}

\usepackage[center]{caption}

\usepackage{tikz}
\usetikzlibrary{matrix,fit}
\usetikzlibrary{shapes,arrows}

\usepackage{ graphicx,  bm,  bbding,pifont}
\usepackage{epsf, epsfig}
\usepackage{graphics}
\usepackage{times,relsize}
\usepackage{fancybox,calc}
\usepackage{multirow}
\usepackage{graphics, color, psfrag,amsfonts}
\usetikzlibrary{decorations.markings}
\usepackage{sidecap}
\usetikzlibrary{shapes,arrows}
\usetikzlibrary{decorations.pathreplacing}
\tikzstyle{decision} = [diamond, draw, fill=white!20,
    text width= 2em, text centered, node distance=1cm, inner sep=2.1pt]
\tikzstyle{block} = [rectangle, draw, fill=white!20,
    text width=2em, text centered, rounded corners, minimum height=2.1em]
\tikzstyle{new_block} = [rectangle, draw, fill=gray!30,
    text width=2em, text centered, rounded corners, minimum height=2.1em]
\tikzstyle{block2} = [rectangle, draw, fill=white!20,
    text width=2em, text centered, rounded corners, minimum height=2.7em]
\tikzstyle{line} = [draw, -latex']
\tikzstyle{state}=[circle, draw, fill=white!10,
    text width=2em, text badly centered, inner sep=2.5pt,minimum height=1.5em]
\tikzstyle{cloud} = [ellipse,fill=gray!35, node distance=3cm,
    minimum height=2em]

\ifCLASSOPTIONonecolumn

\else
    
\fi

\begin{document}
\title{Instantly Decodable Network Coding for Real-Time  Scalable Video Broadcast over Wireless Networks}
\author{\authorblockN{Mohammad S. Karim, ˜\IEEEmembership{Student Member, ˜IEEE,
}  Parastoo Sadeghi, ˜\IEEEmembership{Senior Member, ˜IEEE,
}  Sameh Sorour, ˜\IEEEmembership{Member, ˜IEEE,
} and Neda Aboutorab, ˜\IEEEmembership{Member, ˜IEEE
}}
\thanks{M. S. Karim, P. Sadeghi and N. Aboutorab  are with the  Australian National University,
Australia  (\{mohammad.karim, parastoo.sadeghi, neda.aboutorab\}@anu.edu.au). S. Sorour is with the King Fahd University of Petroleum and Minerals, Saudi Arabia (samehsorour@kfupm.edu.sa).}
}

\maketitle
\ifCLASSOPTIONonecolumn
\vspace{-5mm}
\fi

\vspace{-5mm}
\begin{abstract}
In this paper, we study  a real-time  scalable video broadcast  over wireless  networks in instantly decodable network coded (IDNC) systems. Such  real-time scalable video has a hard deadline  and  imposes a decoding order on the video layers. We first derive the upper bound on the probability that the individual completion times of all receivers  meet the deadline. Using this probability, we design two  prioritized IDNC algorithms, namely the expanding window IDNC (EW-IDNC) algorithm and the non-overlapping window IDNC (NOW-IDNC) algorithm. These algorithms  provide a high level of protection to the most important video layer before considering additional video layers in coding decisions. Moreover, in these algorithms,   we  select an appropriate  packet combination over a given number of video layers  so that  these video layers are decoded by the maximum number of receivers  before the deadline. We formulate this  packet selection problem  as a  two-stage maximal clique selection problem  over an IDNC graph.  Simulation  results  over a real scalable video stream show that  our proposed  EW-IDNC and NOW-IDNC algorithms improve the received video quality compared to the existing IDNC algorithms.
\end{abstract}
\ifCLASSOPTIONonecolumn
\vspace{-1mm}
\fi
\begin{IEEEkeywords}
 Wireless Broadcast, Real-time Scalable Video,  Individual Completion Time, Instantly Decodable Network Coding.
\end{IEEEkeywords}
\ifCLASSOPTIONonecolumn
\vspace{-1mm}
\fi
\section{Introduction} \label{introduction}
Network coding  has shown great potential to improve  throughput,   delay  and quality of services  in  wireless  networks  \cite{katti2006xors,zeng2012joint,wang2014coding,muhammad2014simulation,dong2013delay,yan2011weakly,el2011interplay,li2011capacity,li2011optimal,magli2012network,seferoglu2009video,seferoglu2011cooperative,el2007minimum}. These merits of network coding make it an attractive  candidate  for multimedia applications \cite{magli2012network,seferoglu2009video,seferoglu2011cooperative}. In  this paper, we are interested in real-time  scalable video applications \cite{seeling2004network,schwarz2007overview}, which compress video frames in the form of one \emph{base layer} and several \emph{enhancement layers}.
The base layer provides the basic video quality and  the enhancement
layers provide successive improved video qualities. Using such a scalable video  stream,   the sender  adapts a video bit rate to the available network bandwidth by sending the base layer and as many    enhancement layers as possible. Moreover, the real-time scalable video   has two distinct characteristics. First,  it has  a hard deadline such that the video layers need to be decoded on-time to be usable at the applications. Second,  the video layers exhibit a hierarchical order such that a video  layer can be decoded only if this layer and all its lower layers are  received. Even though scalable video  can tolerate the loss of one or more enhancement  layers, this adversely affects the video quality experienced by viewers. Therefore,  it is desirable to design network coding schemes so that the received  packets before  the deadline contribute to decoding the maximum number of   video layers.

Network coding  schemes  are often  adopted and designed  to be suitable for different  applications. For example,  the works in \cite{vukobratovic2012unequal,wang2012efficient,esmaeilzadeh2014inter,lima2010secure}  adopted random linear network coding (RLNC) strategies  for   scalable video transmission and designed window based RLNC such that  coded packets are formed across all packets in  different numbers of  video layers. In particular, the authors  in  \cite{vukobratovic2012unequal} used a probabilistic approach for selecting coding windows and included the packets in the  lower video layers  into all coded packets  to obtain  high decoding probabilities for the lower  layers.  However, the authors in \cite{esmaeilzadeh2014inter} considered  a  scalable video transmission with a hard deadline and used a  deterministic approach for selecting  coding windows over all  transmissions before the deadline.
Despite the best    throughput performance of  RLNC, in this paper,  we adopt   \emph{instantly decodable network coding} (IDNC) strategies  due to its several attractive properties \cite{li2011capacity,li2011optimal,sadeghi2010optimal,sorour2010decoding,keller2008online,le2013instantly, sorour2010completion,aboutorabenabling,zhan2011coding}. IDNC aims to provide instant packet decodability upon successful packet reception at the receivers.  This instant decodability  property allows a  progressive  recovery of the video layers as the receivers decode more packets. Furthermore, the encoding  process of IDNC is performed using simple XOR operations compared to more    complicated operations over large Galois fields performed in RLNC.
The  XOR operations  also reduce packet overhead compared to    the  coefficient reporting overhead required in RLNC.  The decoding process of IDNC is  performed using  XOR operations, which is  suitable for implementation in simple and cost-efficient receivers, compared to  complex matrix inversion performed  in RLNC.

Due to these attractive properties, the authors in  \cite{sadeghi2010optimal,sorour2010decoding,keller2008online,le2013instantly} considered IDNC for wireless broadcast of a set of packets and aimed to service the maximum number of receivers with a new packet in each transmission. In  \cite{sorour2010completion,aboutorabenabling},  the authors addressed the problem of  minimizing  the number of   transmissions required   for   broadcasting a set of packets  in IDNC systems and formulated the problem into a stochastic shortest path (SSP) framework.  However, the works in \cite{sadeghi2010optimal,sorour2010decoding,keller2008online,le2013instantly,sorour2010completion,aboutorabenabling} neither considered  dependency between  source packets to use at the applications nor considered  explicit packet delivery  deadline. Several other works in IDNC considered  different importance of packets and prioritized packets differently in  coding decisions. In particular, the authors in  \cite{li2011capacity} adopted IDNC for video streaming and showed that their proposed IDNC schemes are asymptotically throughput optimal for the  three-receivers system subject to sequential packet delivery deadline constraints.  However, the work in \cite{li2011capacity} neither considered   dependency between  source packets as is present in the scalable video applications nor considered  an arbitrary number of receivers.  Another work in \cite{seferoglu2009video}  considered a single layer video transmission and  determined the importance of each video packet  based on its  contribution to the video quality. The selected IDNC packet in \cite{seferoglu2009video}  maximized    the  video quality  in the current transmission  without taking into account the coding opportunities and the  video quality over  the successor transmissions before the deadline.

In the context of IDNC for scalable video with multiple layers, the most related works to ours are \cite{muhammad2013instantly,wanginstantly}.
In \cite{muhammad2013instantly}, the authors considered   that a set of packets forming the base layer has high priority compared to an  another set of  packets forming the enhancement layers. However, the IDNC algorithms in \cite{muhammad2013instantly} aimed to  reduce  the number of transmissions required for delivering all the packets instead of giving  priority to  reducing   the number of transmissions required for delivering the high priority packets. The coding decisions  in \cite{muhammad2013instantly} also searched for the existence of a special IDNC packet   that   can  simultaneously reduce the number of transmissions required for delivering the high priority packets and the number of transmissions required for delivering all the packets. On the other hand, the authors in \cite{wanginstantly} discussed the hierarchical order of video layers with motivating examples and  proposed a heuristic packet selection algorithm. The IDNC algorithm in \cite{wanginstantly} aimed to balance between  the number of transmissions required for delivering the base layer and  the  number of transmissions required for delivering all  video layers. Both works in \cite{muhammad2013instantly,wanginstantly} ignored  the hard deadline and did not strictly  prioritize to  deliver the base layer packets before the deadline. However, for real-time scalable video transmission, addressing  the hard deadline for the base layer packets  is essential as all other packets depend on the base layer packets.

In this paper, inspired by  real-time scalable video that has  a  hard deadline and decoding dependency between  video layers, we are interested in designing an efficient IDNC framework that  maximizes the minimum  number of decoded  video layers over all receivers before the deadline (i.e., improves  fairness in terms of the minimum video quality across all receivers).
In such scenarios,  by taking into account the deadline, IDNC schemes  need to make coding decisions over the  packets in the first video layer or the packets   in all video layers.  While the former   guarantees the highest level of protection to the first  video layer, the latter   increases the  possibility of  decoding a large number  of  video layers before the deadline.  In this context,  our main contributions are summarized as follows.
\begin{itemize}
\item We   derive the upper bound on the probability that the individual completion times of all receivers for a given number of video layers meet the deadline. Using this probability, we are able to approximately  determine  whether the broadcast of any given  number of video layers can be completed before the deadline with a  predefined probability.  

\item We  design two prioritized  IDNC algorithms for scalable video, namely the expanding window IDNC (EW-IDNC) algorithm and the non-overlapping window IDNC (NOW-IDNC) algorithm. EW-IDNC algorithm   selects a packet combination over the first  video layer and  computes the resulting upper bound on the probability that  the broadcast of that video layer can be completed before the deadline. Only when this probability meets a predefined high threshold,  the algorithm considers  additional   video layers in  coding decisions  in order  to increase the number of decoded video layers at the receivers.

\item In EW-IDNC and NOW-IDNC algorithms, we select an appropriate packet combination over a given number of video layers  that increases the possibility of decoding those video layers  by the maximum number of receivers before the deadline. We formulate this  problem    as a  two-stage maximal clique selection problem over an IDNC graph. However, the formulated maximal clique selection problem is  NP-hard and even hard to approximate. Therefore,  we exploit the properties of the problem formulation and  design a computationally simple heuristic packet selection algorithm.

\item We use  a  real  scalable video stream to evaluate the performance of our proposed  algorithms. Simulation results  show  that our proposed EW-IDNC and NOW-IDNC algorithms increase the minimum number of decoded  video layers over all receivers compared to  the  IDNC algorithms in \cite{le2013instantly,wanginstantly} and achieve a similar performance compared to the expanding window RLNC  algorithm in  \cite{vukobratovic2012unequal,esmaeilzadeh2014inter} while preserving the benefits of IDNC strategies.
\end{itemize}

The rest of this paper is organized as follows. The system model and  IDNC graph are described in Section~\ref{tools}.  We illustrate the importance of appropriately choosing a  coding window in  Section \ref{window} and  draw several guidelines for prioritized IDNC algorithms in  Section \ref{guidelines}. Using these guidelines, we design two prioritized IDNC algorithms  in Section \ref{scalable}.  We formulate the problem of finding an appropriate packet  combination in Section~\ref{formulation} and design a  heuristic packet selection  algorithm in Section~\ref{heuristic}. Simulation results are presented   in  Section \ref{results}.  Finally,  Section \ref{conclusion} concludes the paper.

\ifCLASSOPTIONonecolumn
\vspace{-1mm}
\fi
\section{Scalable Video Broadcast System}\label{tools}

\subsection{Scalable Video Coding}
We consider a system that employs the scalable video codec (SVC)   extension to H.264/AVC video compression standard \cite{seeling2004network,schwarz2007overview}.  A group of pictures (GOP) in scalable video   has several video layers  and the information bits of each video  layer  is divided
into   one or more   packets.  The video layers  exhibit a hierarchical order  such that each video layer  can only be decoded after successfully receiving all the packets of this layer and its  lower layers.  The first video layer (known as the \emph{base layer}) encodes the lowest temporal, spatial, and quality levels of the original video and the successor video layers  (known as the \emph{enhancement layers}) encode the difference between the video layers of higher temporal, spatial, and quality levels and the base  layer.  With the increase in the  number of decoded video  layers, the  video quality improves at the receivers.

\newtheorem{examples}{\textbf{Example}}

\subsection{System Model}
We consider  a wireless  sender (e.g., a base station  or a wireless access point)  that wants to broadcast a set of $N$ source packets forming a GOP, $\mathcal{N} = \{P_1,...,P_N\}$, to a set  of $M$ receivers, $\mathcal{M} = \{R_1,...,R_M\}$.\footnotemark \footnotetext{Throughout the paper, we use calligraphic letters to denote sets  and  their corresponding capital letters to denote the cardinalities of these sets (e.g., $N = |\mathcal N|$).} A network coding scheme is applied   on the packets of a single GOP as soon as all the packets are  ready, which implies that neither merging of GOPs nor buffering of packets in more than one GOP at the sender is allowed. This significant aspect arises  from  the minimum delivery delay requirement in real-time video streaming. Time is slotted and the sender can transmit one packet per  time slot $t$. There is a limit on the total number of allowable time slots $\Theta$ used to  broadcast the $N$ packets to the $M$ receivers, as the deadline for the current GOP expires after $\Theta$ time slots. Therefore, at any time slot $t \in [1,2,...,\Theta]$, the sender can compute the  number of remaining  transmissions for the current GOP  as, $Q = \Theta-t+1$.


In the scalable video broadcast system,  the sender has  $L$ scalable video layers and each video layer consists of one or more packets. Let the set   $\mathcal{N} = \{P_1^1,P_2^1,...,P_{n_1}^1,...,P_1^L,P_2^L,...,P_{n_L}^L\}$ denote all the packets in the $L$ video layers, with $n_{\ell}$ being the number of packets in the $\ell$-th video layer. In fact, $N = \sum_{\ell=1}^L  n_{\ell}$. Although the number of video layers in a  GOP of a video stream is fixed, depending on the video content, $n_{\ell}$ and $N$ can have different values for different GOPs. We  denote the set that contains  all  packets   in the first   $\ell$ video  layers as $\mathcal{N}^{1:\ell}$ and the cardinality of  $\mathcal{N}^{1:\ell}$ as $N^{1:\ell}$. 

The receivers are assumed to be heterogeneous (i.e., the channels between  the sender and the receivers are not necessarily identical) and each transmitted packet is subject to an independent Bernoulli erasure at receiver $R_i$ with  probability $\epsilon_i$.  Each receiver listens to all transmitted packets  and  feeds back to the sender a positive or negative acknowledgement (ACK or NAK) for each received or lost packet.
After each transmission,  the sender stores the reception  status of all packets at all receivers in an $ M\times N $  \emph{state feedback matrix (SFM)} $\mathbf{F} = [f_{i,j}],$ $\; \forall R_i  \in \mathcal{M}, P_j\in \mathcal{N}$ such that:
 \begin{equation}
  f_{i,j} =
   \begin{cases}
    0 & \text{if packet $P_j$ is received by receiver $R_i$}, \\
    1 & \text{if packet $P_j$ is missing at receiver $R_i$}.
   \end{cases}
 \end{equation}

\newtheorem{remark}{\textbf{Remark}}

\begin{examples}
An example of SFM with $M = 2$ receivers and $N = 5$ packets is given as follows:
\begin{equation}\label{example1}
\mathbf{F} = \begin{pmatrix}
  1 & 0 & 1 & 1 & 1\\
  0 & 1 & 1 & 0 & 0\\
\end{pmatrix}.
\end{equation}
In this example, we assume that  packets $P_1$ and $P_2$  belong to the first (i.e., base) layer,  packets $P_3$ and $P_4$ belong to the second  layer and packet $P_5$ belongs to the third  layer. Therefore, the set containing   all  packets   in the first   two video  layers is $\mathcal{N}^{1:2} =\{P_1,P_2,P_3,P_4\}$.
\end{examples}

\begin{figure}[t]
        \centering
        \includegraphics[width=8cm,height=5cm]{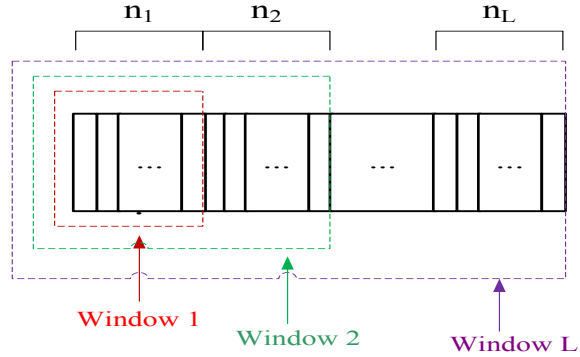}
        \caption{$L$  windows for an $L$-layer GOP with $n_{\ell}$ packets in the $\ell$-th layer.} \label{fig:window}
\end{figure}

\newtheorem{definitions}{\textbf{Definition}}
\begin{definitions}
\emph{A  window over the first $\ell$  video layers (denoted by $\omega_{\ell}$) includes all  the  packets in   $ \mathcal{N}^{1:\ell} = \{P_1^{1},P_2^{1},...,P_{n_{1}}^{1},...,P_1^{\ell},P_2^{\ell},...,P_{n_{\ell}}^{\ell} \}$.}
\end{definitions}
There are $L$  windows  for a GOP  with  $L$ video layers as shown in Fig. \ref{fig:window}. The SFM corresponding to  the window $\omega_{\ell}$ over the first $\ell$ video layers   is an $M\times N^{1:\ell}$ matrix $\mathbf{F}^{1:\ell}$, which contains the first $N^{1:\ell}$ columns of SFM $\mathbf{F}$.

Based on the SFM, the  following two sets of packets can be attributed to each receiver $R_i$   at any given time slot $t$:
\begin{itemize}

\item The \emph{Has set} of receiver $R_i$ in the first  $\ell$ video layers ($\mathcal{H}_i^{1:\ell}$) is defined as the set of packets  that are  decoded by receiver $R_i$ from the first $\ell$ video layers. In Example 1, the Has set of receiver $R_2$ in the first two video layers  is $\mathcal{H}_2^{1:2} = \{P_1, P_4\}$.

\item The \emph{Wants set} of receiver $R_i$ in the first  $\ell$ video layers ($\mathcal{W}_i^{1:\ell}$) is defined as the set of packets  that are missing at  receiver $R_i$  from the first $\ell$ video  layers. In other words, $\mathcal{W}_i^{1:\ell} = \mathcal{N}^{1:\ell} \setminus \mathcal{H}_i^{1:\ell}$.  In Example 1, the Wants set   of receiver $R_2$ in the first two video layers  is $\mathcal{W}_2^{1:2} = \{P_2, P_3\}$.

\end{itemize}
The cardinalities of $\mathcal{H}_i^{1:\ell}$  and $ \mathcal{W}_i^{1:\ell}$  are denoted  by $H_i^{1:\ell}$ and $W_i^{1:\ell}$, respectively. The set of receivers having \emph{non-empty Wants sets} in the first  $\ell$ video  layers is denoted by $\mathcal M_w^{1:\ell}$ (i.e., $\mathcal{M}_w^{1:\ell} = \left\{R_i \middle | \mathcal{W}_i^{1:\ell} \neq \varnothing \right\}$).
At any given SFM $\mathbf F^{1:\ell}$ at time slot $t$,  receiver $R_i$ having non-empty Wants set  in the first  $\ell$ video layers (i.e., $R_i \in \mathcal{M}_w^{1:\ell}$)  belongs to one of the following  three sets:

\begin{itemize}
\item The \emph{critical set} of receivers for the first  $\ell$ video layers ($\mathcal{C}^{1:\ell}$) is defined as   the set of receivers with the number of  missing packets in the first  $\ell$  video layers  being equal to the number of remaining  $Q$ transmissions (i.e., $W_i^{1:\ell} = Q, \forall R_i \in \mathcal{C}^{1:\ell}$).

 \item The \emph{affected set} of receivers for the first  $\ell$ video layers ($\mathcal{A}^{1:\ell}$) is defined as the set of receivers with the number of  missing packets in the first   $\ell$ video  layers being greater than  the number of remaining $Q$ transmissions  (i.e., $W_i^{1:\ell} > Q, \forall R_i \in \mathcal{A}^{1:\ell}$).

 \item The \emph{non-critical set} of receivers for the first  $\ell$ video layers ($\mathcal{B}^{1:\ell}$) is defined as the set of receivers with the number of  missing packets in the first   $\ell$  video layers  being less than the number of remaining $Q$  transmissions (i.e., $W_i^{1:\ell} < Q, \forall R_i \in \mathcal{B}^{1:\ell}$).
\end{itemize}

In fact, $\mathcal{C}^{1:\ell} \cup \mathcal{A}^{1:\ell} \cup \mathcal{B}^{1:\ell} = \mathcal{M}_w^{1:\ell}$. We denote the cardinalities of $\mathcal{C}^{1:\ell}$, $\mathcal{A}^{1:\ell}$ and $\mathcal{B}^{1:\ell}$ as $C^{1:\ell}$, $A^{1:\ell}$ and $B^{1:\ell}$, respectively.


\begin{definitions}
\emph{A transmitted packet is instantly decodable for receiver $R_i$ if it contains exactly one  source packet from $\mathcal{W}_i^{1:L}$.}
\end{definitions}
\begin{definitions}
\emph{Receiver $R_i$ is targeted by  packet $P_j $ in a  transmission  when this receiver  will immediately decode  missing packet $P_j$ upon  successfully receiving   the transmitted packet.}
\end{definitions}
\begin{definitions}
\emph{At   time slot $t$, individual completion time  of  receiver $R_i$ for the first  $\ell$ video  layers  (denoted by $T_{W_i^{1:\ell}}$) is  the total  number of transmissions required to deliver all the missing  packets in $\mathcal{W}_i^{1:\ell}$ to   receiver $R_i$.}
\end{definitions}
Individual completion time of receiver $R_i$ for the first $\ell$ video  layers  can be $T_{W_i^{1:\ell}} = W_i^{1:\ell}, W_i^{1:\ell}+1,...$ depending on the number of transmissions that receiver $R_i$ is targeted with a new packet and the  channel erasures experienced by receiver $R_i$ in those transmissions.

\begin{definitions}
\emph{At time slot $t$,  individual completion times  of  all receivers  for the first  $\ell$ video  layers  (denoted by $T^{1:\ell}$) is the total number of transmissions required to deliver all the missing packets from the first  $\ell$ video  layers  to all receivers in  $\mathcal M_w^{1:\ell}$. }
\end{definitions}
In other words,   given SFM $\mathbf F^{1:\ell}$ at time slot $t$, $T^{1:\ell}$ defines  the total number of transmissions required to complete the broadcast of $\ell$ video layers.

\begin{definitions}
\emph{At time slot $t$,  individual completion times  of  all non-critical receivers  for the first  $\ell$ video  layers  (denoted by $T_{B}^{1:\ell}$) is the total number of transmissions required to deliver all the missing packets from the first  $\ell$ video  layers  to all non-critical receivers in  $\mathcal B^{1:\ell}$. }
\end{definitions}

\subsection{IDNC Graph and  Packet Generation}\label{sec:IDNCgraph}

We define the representation of all  feasible packet combinations that are instantly decodable by a subset of, or  all  receivers,  in the form of a graph. As described in \cite{sorour2010decoding,sorour2010completion}, the IDNC graph $\mathcal G(\mathcal V, \mathcal E)$ is constructed  by first inducing a vertex $v_{ij} \in \mathcal V$ for each missing  packet $P_j\in \mathcal{W}_i^{1:L}, \; \forall R_i\in \mathcal{M}$. Two vertices $v_{ij}$ and $v_{mn}$ in $\mathcal{G}$  are connected (adjacent)  by an edge $e_{ij,mn}\in \mathcal E$, when one of the following two conditions holds:
\begin{itemize}
 \item \textbf{C1}: $P_j = P_n$, the two vertices are induced by the same missing  packet $P_j$ of two different receivers $R_i$ and $R_m$. 
 \item \textbf{C2}: $P_j\in \mathcal{H}_m^{1:L}$ and $P_n\in \mathcal{H}_i^{1:L}$, the requested packet of each vertex is in the Has set of the receiver of the other vertex. 
 \end{itemize}

\begin{table}
 \caption{Main notations  and their descriptions} \label{table:sum}
 \centering
    \begin{tabular}{|p{0.8cm}|p{6.4cm} |p{0.8cm}  |p{6.4cm} |}
    \hline
    \textit{} &  \textit{Description} &  \textit{} &  \textit{Description} \\ \hline
    $Q$ & Number of remaining transmissions   &  $L$  & Number of video layers \\ \hline
    $\mathcal{M}$ & Set of $M$ receivers  &  $\mathcal{N}^{1:\ell}$  & Set of $N^{1:\ell}$ packets  \\ \hline
    $R_i$ & The $i-$th receiver in $\mathcal{M}$  &  $P_j$  &The $j-$th  packet in $\mathcal{N}$ \\ \hline
    $\mathbf{F}^{1:\ell}$ & $M \times N^{1:\ell}$ state feedback matrix  &  $\omega_{\ell}$ & $\ell$-th window among $L$  windows \\ \hline
    $\mathcal{H}_i^{1:\ell}$ & Has set  of receiver $R_i$ in  $\ell$  layers &  $\mathcal{W}_i^{1:\ell}$  & Wants set  of receiver $R_i$ in  $\ell$  layers \\ \hline
    $T_{W_i^{1:\ell}}$  & Individual completion time of receiver $R_i$ for  $\ell$  layers  &  $T^{1:\ell}$ & Individual completion times of all receivers  for  $\ell$  layers  \\ \hline
    $ \mathcal G^{1:\ell} $  & IDNC graph constructed from  $\mathbf F^{1:\ell}$ &  $T_B^{1:\ell}$ & Individual completion times of all  non-critical    receivers for $\ell$ layers \\ \hline
    $\mathcal{C}^{1:\ell}$ & Critical set of receivers for   the first $\ell$  layers &  $\mathcal{A}^{1:\ell}$ & Affected set of receivers  for  the first $\ell$  layers   \\ \hline
    $\mathcal{B}^{1:\ell}$ & Non-critical set of receivers  for  the first $\ell$  layers &  $\mathcal{M}_w^{1:\ell}$ & Set of receivers having non-empty Wants sets  in the first $\ell$  layers   \\ \hline
    $v_{ij}$ & A vertex in an IDNC graph induced by missing packet $P_j$ at receiver $R_i$ &   $\mathcal{X}(\kappa)$  & Set of targeted receivers in maximal clique $\kappa$ \\ \hline
    \end{tabular}
\end{table}

Given this graph representation, the set of all feasible IDNC packets   can be defined by the set of all maximal cliques in   graph $\mathcal{G}$.\footnotemark \footnotetext{In an undirected graph, all vertices in a clique are connected to each other with  edges. A clique is maximal if it is not a subset of any larger clique \cite{garey1979computers}.} The sender can generate an IDNC  packet for a given transmission by XORing all the source packets identified by the vertices of a selected maximal clique (represented  by $\kappa$)  in  graph $\mathcal G$. Note that each receiver can have at most one vertex (i.e., one missing packet) in a maximal clique  $\kappa$  and the selection of a maximal clique $\kappa$ is equivalent to the selection of \emph{a set of targeted receivers} (represented by $\mathcal{X}(\kappa)$).  A summary of the main notations used in this  paper is presented in Table \ref{table:sum}.


\ifCLASSOPTIONonecolumn
\vspace{-1mm}
\fi

\section{Importance  of Appropriately Choosing a Coding Window} \label{window}
In scalable video with multiple layers, the sender needs to  choose a window of video layers and the corresponding SFM to select a packet combination in each transmission. In general,  different  windows   lead  to  different packet combinations  and result in different   probabilities  of completing the broadcast of  different numbers of video layers  before the deadline.  To further illustrate, let us  consider the following  SFM  with $M = 2$ receivers and $N = 2$ packets at time slot $t$:
\begin{equation}\label{example22}
\mathbf{F} = \begin{pmatrix}
   0 & 1 \\
   1 & 1 \\
\end{pmatrix}.
\end{equation}
In this scenario, we assume that packet $P_1$  belongs to the first video layer and  packet $P_2$ belongs to the second video layer. We further assume that  there are  two remaining transmissions before the deadline, i.e., $Q = 2$. Given two video layers, there are two windows such as  $\omega_1 = \{P_1\}$ and  $\omega_2 = \{P_1, P_2\}$. With these windows,  the possible packet transmissions at time slot $t$ are:
\begin{itemize}
\item \textbf{Case 1:} Window $\omega_1$ leads to  packet $P_1$  transmission since  it targets  receiver $R_2$ and $\mathcal M_w^{1:1} = \{R_2\}$.
\item \textbf{Case 2:} Window $\omega_2$ leads to  packet $P_2$ transmission since it  targets  receivers $R_1$ and $R_2$ and $\mathcal M_w^{1:2} = \{R_1, R_2\}$.
\end{itemize}

\textbf{(Case 1:)} With packet $P_1$ transmitted  at time slot $t$, we can compute the probabilities of completing the broadcast of  different numbers of video layers  before the deadline as follows.
\begin{enumerate}
\item The probability of completing the first video layer broadcast before the deadline can be computed as, $\mathds{P} [T^{1:1} \leq 2] = (1-\epsilon_2) + \epsilon_2(1-\epsilon_2)$.
    \begin{itemize}
     \item   $(1-\epsilon_2)$ defines the  packet reception probability at receiver $R_2$ at time slot $t$.
     \item  $\epsilon_2(1-\epsilon_2)$  defines the probability that  packet $P_1$   is lost at receiver $R_2$ at time slot $t$ and  is received at receiver $R_2$  at time slot $t+1$.
     \end{itemize}
\begin{remark}\label{remark:attempt}
\emph{It can be stated that  the  missing packets of all receivers need  to be attempted at least  once in order to have a possibility of delivering  all the missing packets to all receivers.}
\end{remark}
\item Using Remark \ref{remark:attempt}, the sender  transmits packet $P_2$ at time slot $t+1$. Consequently, the probability of completing both video layers' broadcast before the deadline can be computed as,  $\mathds{P} [T^{1:2} \leq 2] = (1-\epsilon_2)(1-\epsilon_1) (1-\epsilon_2)$. This is the probability    that each missing packet  is  received from  one transmission (i.e., one attempt).
\end{enumerate}
A summary of   probability expressions  used throughout Case 1  can be found in Table \ref{T:case1}.

\begin{table}
 \caption{Probability expressions used in Case 1}\label{T:case1}
 \centering
\begin{tabular}{cc|c|c|c|c|c|c|c|c|c|l}
\cline{3-6}
\cline{8-9}
& & $P_1(t)$ & -& $P_1(t)$ & $P_1(t+1)$ & & $P_1(t)$ & $P_2(t+1)$ \\ \cline{1-9}
\multicolumn{1}{ |c  }{\multirow{2}{*}{$\mathds{P}[T^{1:1} \leq 2]$} } &
\multicolumn{1}{ |c| }{$R_1$} &- &- & - & -  & \multicolumn{1}{ |c  }{\multirow{2}{*}{$\mathds{P}[T^{1:2} \leq 2]$} }   & \multicolumn{1}{ |c| }{-} & $(1-\epsilon_1)$    \\ \cline{2-6} \cline{8-9}
\multicolumn{1}{ |c  }{}                        &
\multicolumn{1}{ |c| }{$R_2$} &$(1-\epsilon_2)$ & - &  $\epsilon_2$ & $(1-\epsilon_2)$ & \multicolumn{1}{ |c  }{}   &  \multicolumn{1}{ |c| }{$(1-\epsilon_2)$} & $(1-\epsilon_2)$  \\ \cline{1-9}
\end{tabular}
\end{table}

\begin{table}
 \caption{Probability expressions used in Case 2}\label{T:case2}
 \centering
\begin{tabular}{cc|c|c|c|c|c|c|c|c|l}
\cline{2-3}
\cline{5-8}
&  $P_2(t)$ & $P_1(t+1)$ & & $P_2(t)$ & $P_1\oplus P_2$ & $P_2(t)$ & $ P_1(t+1)$ \\ \cline{1-8}
\multicolumn{1}{ |c  }{}{\multirow{2}{*}{$\mathds{P}[T^{1:1} \leq 2]$} }{} &
\multicolumn{1}{ |c| }{-}  &-  & \multicolumn{1}{ |c  }{\multirow{2}{*}{$\mathds{P}[T^{1:2} \leq 2]$} }   & \multicolumn{1}{ |c| }{$\epsilon_1$ } & $(1-\epsilon_1)$ & $(1-\epsilon_1)$ & -  \\ \cline{2-3} \cline{5-8}
\multicolumn{1}{ |c  }{}                        &
\multicolumn{1}{ |c| }{-} & $(1-\epsilon_2)$ & \multicolumn{1}{ |c  }{}   &  \multicolumn{1}{ |c| }{$(1-\epsilon_2)$ } & $(1-\epsilon_2)$ & $(1-\epsilon_2)$ & $(1-\epsilon_2)$  \\ \cline{1-8}
\end{tabular}
\end{table}

\textbf{(Case 2:)} With packet $P_2$ transmitted  at time slot $t$, we can compute the probabilities of completing the broadcast of  different numbers of video layers before the deadline as follows.
\begin{enumerate}
\item The sender  transmits  packet $P_1$  at time slot $t+1$. Consequently,  the probability of completing the first video layer broadcast before the deadline can be computed as, $\mathds{P} [T^{1:1} \leq 2] = (1-\epsilon_2)$. This is the probability that packet $P_1$  is received  at  receiver $R_2$  at time slot $t+1$.
\item Using Remark \ref{remark:attempt}, the sender transmits either coded packet $P_1 \oplus P_2$ or packet $P_1$ at time slot $t+1$. Consequently, the probability of completing both video layers' broadcast before the deadline can be computed as, $\mathds{P} [T^{1:2} \leq 2] = \epsilon_1(1-\epsilon_2) (1-\epsilon_1)(1-\epsilon_2) + (1-\epsilon_1)(1-\epsilon_2) (1-\epsilon_2)$.
  \begin{itemize}
  \item  $\epsilon_1(1-\epsilon_2)(1-\epsilon_1)(1-\epsilon_2) $ represents coded packet $P_1 \oplus P_2$ transmission at time slot $t+1$.  The transmitted  packet $P_2$ at time slot $t$ can be  lost  at receiver $R_1$ with probability  $\epsilon_1$ and  can be received at receiver $R_2$ with probability $(1-\epsilon_2)$. With this  loss and reception status, the sender  transmits coded packet $P_1 \oplus P_2$ to target both receivers and the probability that both receivers  receive the transmitted  packet  is $ (1-\epsilon_1) (1-\epsilon_2)$.

   \item  $(1-\epsilon_1)(1-\epsilon_2) (1-\epsilon_2)$  represents  packet $P_1$ transmission at time slot $t+1$.  This is the  probability that each missing packet  is  received from one attempt.
   \end{itemize}
\end{enumerate}

A summary of   probability expressions  used throughout Case 2  can be found in Table \ref{T:case2}.
Using  the results in Case 1 and Case 2, for  given time slot $t$, we can conclude  that:
\begin{itemize}
\item Packet $P_1$ transmission resulting from window $\omega_1$  is a better decision in terms of completing  the first  video layer broadcast  since $\mathds{P} [T^{1:1} \leq 2]$ is larger in Case 1.

\item Packet $P_2$ transmission resulting from window $\omega_2$  is a better decision  in terms of completing both video layers  broadcast   since  $\mathds{P} [T^{1:2} \leq 2]$ is larger in Case 2.
\end{itemize}

\begin{remark}
\emph{The above  results illustrate  that it is not always possible to select a  packet combination   that achieves   high  probabilities  of completing the broadcast of different numbers of video layers before the deadline. In general, some packet transmissions (resulting from different windows) can increase the probability of completing the broadcast of the first video layer, but reduce the probability of completing the broadcast of  all  video layers and vice versa. }
\end{remark}
\ifCLASSOPTIONonecolumn
\vspace{-1mm}
\fi

\section{Guidelines for Prioritized  IDNC Algorithms} \label{guidelines}
In this section, we systematically draw several guidelines for the prioritized  IDNC algorithms that can   maximize the minimum number of decoded video layers over all receivers before the deadline.

\subsection{Feasible  Windows of Video Layers}\label{windowS}

With the assist of the  following definitions, for a given SFM $\mathbf{F}$ at  time slot $t$, we  determine the video layers which can be included in a feasible window and  can  be considered in coding decisions.
\begin{definitions}
\emph{The smallest  feasible window  (i.e., window $\omega_{\ell}$) includes  the minimum number of successive  video layers  such that the Wants set of at least one  receiver in those video layers is non-empty. This can be defined as, $\omega_{\ell} = \min\{|\omega_{1}|,...,|\omega_{L}|\}$ such that  $\exists R_i | \mathcal W_{i}^{1:\ell} \neq \varnothing$.}
\end{definitions}
 In this paper, we  address the  problem of maximizing the  minimum number of decoded video layers over all receivers. Therefore, we define the largest feasible window as follows:
\begin{definitions}
\emph{The largest feasible window (i.e., window $\omega_{\ell+\mu}$ , where $\mu$ can be $0,1,...,L-\ell$) includes the maximum number of  successive video layers  such that the Wants sets of all receivers in those video layers are less than or equal to the remaining $Q$ transmissions.  This can be defined as,  $\omega_{\ell+\mu} = \max \{|\omega_{1}|,...,|\omega_{L}|\}$ such that $\mathcal W_{i}^{1:\ell+\mu} \leq Q, \forall R_i \in \mathcal M$.}
\end{definitions}

Note that there is no affected receiver over the largest  feasible window $\omega_{\ell+\mu}$ (i.e., all receivers belong to critical  and non-critical sets in the first $\ell+\mu$ video layers) since an affected receiver will definitely not be able to decode all its  missing  packets within remaining $Q$ transmissions. An exception to considering no affected receiver  in the largest  feasible window is when it is the smallest feasible window, i.e., $\omega_{\ell + \mu} = \omega_{\ell}$, in which case it is possible $\mathcal{A}^{1:\ell} (t) \neq \varnothing$.


\begin{definitions}
\emph{A feasible window includes any number  of successive    video layers  ranging from the smallest feasible window $ \omega_{\ell}$ to the largest feasible  window $\omega_{\ell + \mu}$. In other words,  a feasible window  can be any  window from $\{\omega_{\ell}, \omega_{\ell+1},...,\omega_{\ell + \mu}\}$.}
\end{definitions}
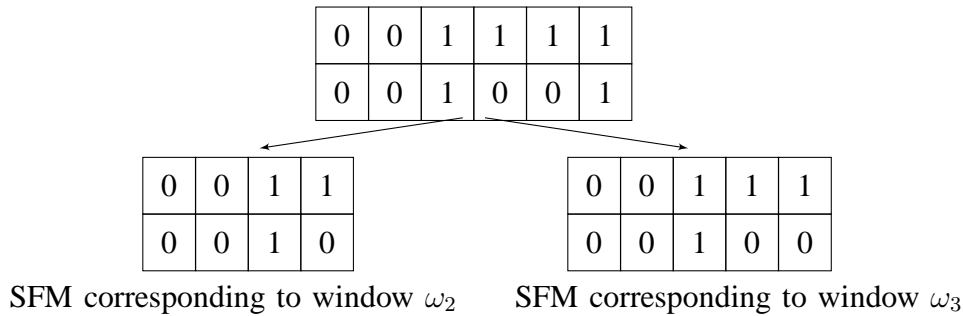
\begin{figure}
\centering
\tikzset{
    table nodes/.style={
        rectangle,
        draw=black,
        align=center,
        minimum height=7mm,
        text depth=0.5ex,
        text height=2ex,
        inner xsep=0pt,
        outer sep=0pt
    },
    table/.style={
        matrix of nodes,
        row sep=-\pgflinewidth,
        column sep=-\pgflinewidth,
        nodes={
            table nodes
        },
        execute at empty cell={\node[draw=none]{};}
    }
}

\begin{tikzpicture}

\matrix (first) [table,text width=7mm,name=table]
{
0 & 0 & 1 & 1 & 1 & 1\\
0 & 0 & 1 & 0 & 0 & 1\\
};

\begin{scope}[xshift=-3 cm, yshift=-2cm]
\matrix (second) [table,text width=7mm,name=table] {
0 & 0 & 1 & 1 \\
0 & 0 & 1 & 0 \\
};
\end{scope}

\begin{scope}[xshift=3cm, yshift=-2cm]
\matrix (second) [table,text width=7mm,name=table] {
0 & 0 & 1 & 1 & 1 \\
0 & 0 & 1 & 0 & 0 \\
};
\end{scope}

\node        (T){};
\node   [below = .55cm]   (T_a){};
\node   [right = -3cm, below = 1cm]   (T_c){};
\node   [right = 3cm, below = 1cm]   (T_d){};
\path [line] (T_a) -- node[name=t2] {}(T_c);
\path [line] (T_a) -- node[name=t3] {} (T_d);

\node  (A_a) [right = -3.2cm, below = 2.1cm] at  (T_a){SFM  corresponding to window $\omega_2$};
\node  (A_b) [right = 3.6cm] at  (A_a){SFM  corresponding to window $\omega_3$};
\end{tikzpicture}
\caption{SFMs corresponding to the feasible windows in Example \ref{ex:windows}} \label{stateTran}
\end{figure}

\begin{examples} \label{ex:windows}
To further illustrate  these feasible windows, consider   the following  SFM at time slot $t$:
\begin{equation}\label{example55}
\mathbf{F} = \begin{pmatrix}
  0 & 0 & 1 & 1 & 1 & 1\\
  0 & 0 & 1 & 0 & 0 & 1\\
\end{pmatrix}.
\end{equation}
In this example, we assume that  packets $P_1$ and $P_2$  belong to the first video layer,  packets $P_3$ and $P_4$ belong to the second video layer, packet $P_5$ belongs to the third video layer and packet $P_6$ belongs to the fourth video  layer. We also assume that  number of remaining transmissions $Q = 3$. The smallest feasible window includes the first two video layers (i.e., $\omega_2 = \{P_1,P_2,P_3,P_4\}$) and the largest feasible window includes the first three video layers (i.e., $\omega_3 = \{P_1,P_2,P_3,P_4,P_5\}$). Note that the fourth video layer is not included in the largest feasible window since receiver $R_1$ has three missing packets in the first three layers, which is already equal to the number of  remaining three transmissions (i.e., $W_1^{1:3}  = Q = 3$). Fig. \ref{stateTran} shows the extracted SFMs from SFM  in \eqref{example55} corresponding  to  the feasible windows.
\end{examples}

\subsection{Probability that the Individual Completion Times  Meet the Deadline} \label{lower bound}

The works in \cite{sorour2010completion,aboutorabenabling} showed  that finding the optimal IDNC schedule  for  minimizing the overall completion time  is computationally intractable due to the curse of dimensionality of dynamic programming. Indeed,  the  random nature of channel erasures  requires the consideration of  all possible  SFMs and  their possible coding decisions  to find the optimal IDNC schedule. With the aim of designing low complexity prioritized  IDNC algorithms, after selecting a packet combination over any given feasible window $\omega_{\ell}$ at  time slot $t$, we compute the resulting upper bound on the  probability that the individual completion times of all receivers for the first $\ell$ video layers  is less than or equal to  remaining $Q-1$ transmissions (represented by $ \hat{\mathds P}^{(t+1)} [T^{1:\ell} \leq Q-1]$ and defined in \eqref{eq:computePP}). Since this probability is computed separately for each receiver and  ignores the interdependence of receivers' packet reception  captured in the SFM, its  computation is simple and does not  suffer from the curse of dimensionality as in \cite{sorour2010completion,aboutorabenabling}.

To derive   probability $ \hat{\mathds P}^{(t+1)} [T^{1:\ell} \leq Q-1]$, we first consider a scenario with one sender and one receiver $R_i$.  Here, individual completion time of this receiver  for the first $\ell$ layers can be $ T_{W_i^{1:\ell}} = W_i^{1:\ell}, W_i^{1:\ell}+1,...$.  The probability of  $T_{W_i^{1:\ell}}$ being equal to $W_i^{1:\ell} + z, z \in [0,1,...,Q-W_i]$  can be expressed using negative binomial distribution as:
\begin{equation} \label{eq:binomial}
  \mathds{P}[T_{W_i^{1:\ell}} = W_i^{1:\ell} + z] =  \binom{W_i^{1:\ell} + z -1}{z} (\epsilon_i)^z(1-\epsilon_i)^{W_i^{1:\ell}}.
\end{equation}
Consequently, the probability that  individual completion time $T_{W_i^{1:\ell}}$ of  receiver $R_i$ is less than or equal to  remaining $Q$ transmissions  can be expressed as:
\begin{equation} \label{eq:metric}
\mathds{P}[T_{W_i^{1:\ell}} \leq Q] = \sum_{z = 0}^{Q - W_i^{1:\ell} } \mathds{P}[T_{W_i^{1:\ell} } = W_i^{1:\ell}  + z].
\end{equation}
We  now consider a scenario with one sender and multiple receivers in $\mathcal M_w^{1:\ell}$. We assume that all  receivers in $\mathcal M_w^{1:\ell}$ are targeted with a new packet in each transmission. This is an  ideal scenario and defines  a lower bound on individual completion time of each receiver. Consequently, we can compute  an upper bound on the probability that  individual completion time of each receiver meets the deadline. Although this ideal scenario is not likely to occur, especially in   systems with large numbers of receivers and packets, we can still use this probability upper bound as a metric in designing our computationally simple  IDNC algorithms.
Having described the ideal scenario with multiple receivers, for a given feasible window $\omega_{\ell}$ at  time slot $t$, we compute  the upper bound on the probability that  individual  completion times of all receivers for the first $\ell$ video layers  is less than or equal to   remaining $Q$ transmissions as:
\begin{equation} \label{eq:metric}
\hat{\mathds{P}}^{(t)}[T^{1:\ell}\leq Q] = \prod_{R_i \in \mathcal{M}_w^{1:\ell} } \sum_{z=0}^{Q - W_i^{1:\ell}} \mathds{P}[T_{W_i^{1:\ell}} = W_i^{1:\ell} + z].
\end{equation}

Due to the instant decodability constraint, it may not be possible to target  all receivers in $\mathcal M_w^{1:\ell}$ with a new packet   at time slot $t$.  After selecting a packet combination over a given feasible window $\omega_{\ell}$ at  time slot $t$, let $\mathcal X$ be the set of targeted receivers and $\mathcal{M}_w^{1:\ell} \setminus \mathcal X$ be the set of ignored receivers. We can express  the resulting upper bound on the probability that the individual  completion times of all  receivers for the first $\ell$ video layers, starting from the successor time slot $t+1$, is less than or equal to the remaining $Q-1$ transmissions as:

\begin{align} \label{eqn:formulation3342}
\hat{\mathds{P}}^{(t+1)} [T^{1:\ell}\leq Q-1]  &=  \prod_{R_i \in \mathcal{X}} \left( \mathds{P}^{(t)} [T_{W_i^{1:\ell}-1} \leq Q-1].(1-\epsilon_i)+ \mathds{P}^{(t)} [T_{W_i^{1:\ell}} \leq Q-1].(\epsilon_i) \right) \nonumber\\
              & \prod_{R_i \in \mathcal{M}_w^{1:\ell} \setminus \mathcal X}  \mathds{P}^{(t)} [T_{W_i^{1:\ell}} \leq Q-1].
\end{align}
\begin{itemize}
\item In the first product in expression  \eqref{eqn:formulation3342}, we compute the probability that a targeted  receiver  receives  its $W_i^{1:\ell}-1$ or $W_i^{1:\ell}$ missing packets in the remaining $Q-1$ transmissions. Note that the number of missing packets at a targeted  receiver  can be  $W_i^{1:\ell} -1$  with its packet reception probability $(1-\epsilon_i)$ or  can be  $W_i^{1:\ell}$ with its channel erasure probability $\epsilon_i$.

  \item In the  second product in expression  \eqref{eqn:formulation3342}, we compute the probability that an ignored  receiver  receives  its  $W_i^{1:\ell}$ missing packets in the remaining $Q-1$ transmissions.
\end{itemize}
By taking  expectation of packet reception and loss cases in the first product in  \eqref{eqn:formulation3342}, we can simplify expression  \eqref{eqn:formulation3342}   as:
\begin{align} \label{eq:metric22}
\hat{\mathds{P}}^{(t+1)} [T^{1:\ell}\leq Q-1] = \prod_{R_i \in \mathcal{X}}  \mathds{P}^{(t)} [T_{W_i^{1:\ell}} \leq Q] \prod_{R_i \in \mathcal{M}_{w}^{1:\ell} \setminus  \mathcal{X}}  \mathds{P}^{(t)} [T_{W_i^{1:\ell}} \leq Q-1].
\end{align}
Note that  a critical and ignored  receiver $R_i \in \{\mathcal{C}^{1:\ell} \cap (  \mathcal{M}_{w}^{1:\ell} \setminus  \mathcal{X})\}$ cannot decode  all missing packets in $W_i^{1:\ell}$   in the remaining $Q-1$ transmissions  since $W_i^{1:\ell}$ is   equal to $Q$ transmissions for a critical receiver.  With this and an exceptional case of having affected receivers described  in Section \ref{windowS}, we can set:
\begin{equation} \label{eq:computePP}
\begin{split}
 \hat{\mathds{P}}^{(t+1)} &[T^{1:\ell} \leq Q-1]\\
 &=   \begin{cases}
    0 & {} \\
    \;\;\;\;\;\;\;\;\;\;\;\;\;\;\;\;\;\;\;\;\;\;\;\;\;\;\;\;\;\;\;\;\;\;\;\;\;\;\;\;\;\;\;\;\;\;\;\;
     \text{If $\mathcal{C}^{1:\ell} \cap (\mathcal{M}_{w}^{1:\ell} \setminus  \mathcal{X})\neq \varnothing$ or $\mathcal{A}^{1:\ell}  \neq \varnothing$} &{} \\
    \prod_{R_i \in \mathcal{X}}  \mathds{P}^{(t)} [T_{W_i^{1:\ell}} \leq Q] \prod_{R_i \in \mathcal{M}_{w}^{1:\ell} \setminus  \mathcal{X}}  \mathds{P}^{(t)} [T_{W_i^{1:\ell}} \leq Q-1]  & \text{Otherwise}
   \end{cases}
   \end{split}
 \end{equation}
In  this paper, we  use expression  \eqref{eq:computePP} as a metric in designing the  computationally simple IDNC algorithms for real-time scalable video.

\subsection{Design Criterion for Prioritized IDNC Algorithms} \label{sec:dc}
In Section \ref{window}, we showed that  some windows and subsequent packet transmissions increase the probability of completing the broadcast of the first  video layer, but reduce the probability of
completing the broadcast of all video layers and vice versa. This complicated interplay of selecting an
appropriate window motivates us to define a design criterion. The objective of the  design criterion is to   expand the coding window over the successor  video layers  (resulting in an increased possibility of completing the broadcast of  those video layers)  after providing  a certain level of protection to  the lower  video   layers.

\newtheorem{designs}{\textbf{Design Criterion}}
\begin{designs}
\emph{The design criterion for the first  $\ell$ video layers   is   defined as the probability $\hat{\mathds{P}}^{(t+1)} [T^{1:\ell} \leq Q-1]$ meets a certain threshold $\lambda$  after selecting a packet combination  at time slot $t$.}
\end{designs}
In other words, the design criterion for the first  $\ell$ video layers  is satisfied when logical condition $\hat{\mathds{P}}^{(t+1)} [T^{1:\ell} \leq Q-1] \geq \lambda$ is true after selecting a packet combination at time slot $t$. Here, probability $\hat{\mathds{P}}^{(t+1)} [T^{1:\ell} \leq Q-1]$ is computed using  expression  \eqref{eq:computePP} and  threshold  $\lambda$  is chosen in a deterministic manner according to the level  of  protection  provided to each  video layer.

In scalable video applications, each decoded  layer  contributes to the video quality and the  layers are decoded following the hierarchical order. Therefore,  the selected packet combination at time slot $t$ requires  to satisfy the design criterion following the decoding order of the video layers. In other words, the first priority is satisfying the design criterion for the first   video layer (i.e., $\hat{\mathds{P}}^{(t+1)} [T^{1:1} \leq Q-1] \geq \lambda$), the second priority is satisfying the design criterion for the first two video layers (i.e., $\hat{\mathds{P}}^{(t+1)} [T^{1:2} \leq Q-1] \geq \lambda$) and so on. Having satisfied such a prioritized design criterion,  the coding window  can continue to expand over the successor video layers to increase the possibility of completing the broadcast of those video layers.


\section{Prioritized  IDNC Algorithms for Scalable Video}\label{scalable}
In this section, using the guidelines drawn in Section \ref{guidelines},  we  design two prioritized  IDNC  algorithms that  increase the probability  of completing the broadcast of a large number of video layers before the deadline. These algorithms  provide unequal levels of  protection to the video layers and adopt prioritized  IDNC strategies  to meet  the hard deadline for the most important video layer in each transmission.

\subsection{Expanding Window Instantly Decodable Network Coding  (EW-IDNC) Algorithm}
Our proposed \emph{expanding window instantly decodable network coding} (EW-IDNC) algorithm  starts by selecting a packet combination over the smallest feasible  window and  iterates by selecting a new packet combination over each   expanded feasible  window while satisfying the   design criterion   for  the video layers in each window. Moreover, in EW-IDNC algorithm,  a packet combination (i.e., a maximal clique $\kappa$)   over a given feasible window is selected following   Section \ref{formulation} or Section  \ref{heuristic}. 


At \emph{ Step 1 of Iteration 1},  the EW-IDNC algorithm  selects   a  maximal clique $\kappa$  over the smallest feasible window  $\omega_{\ell}$.    At \emph{Step 2 of Iteration 1}, the  algorithm computes the probability  $\hat{\mathds{P}}^{(t+1)} [T^{1:\ell} \leq Q-1]$ using expression  \eqref{eq:computePP}.   At \emph{Step 3 of Iteration 1}, the  algorithm  performs one of the following two steps.
\begin{itemize}
\item  It  proceeds to Iteration 2 and considers window $\omega_{\ell+1}$, if $\hat{\mathds{P}}^{(t+1)} [T^{1:\ell} \leq Q-1] \geq \lambda$ and $|\omega_{\ell}| <  |\omega_{\ell + \mu}|$. This is  the case when the design criterion for the first $\ell$ video layers is satisfied and the  window can be further expanded.
 \item It   broadcasts the selected $\kappa$ at this Iteration 1, if $\hat{\mathds{P}}^{(t+1)} [T^{1:\ell} \leq Q-1] < \lambda$ or $|\omega_{\ell}| =  |\omega_{\ell + \mu}|$. This is  the case when the design criterion for the first $\ell$ video layers is not satisfied or the  window is already the largest feasible window.
\end{itemize}

\IncMargin{1em}
\begin{algorithm}[!t]
\textbf{(Iteration 1)} Consider the smallest feasible window  $\omega_{\ell}$\;
      Select maximal clique $\kappa$   over window $\omega_{\ell}$\;
       Compute probability $\hat{\mathds{P}}^{(t+1)} [T^{1: \ell } \leq Q-1]$ using expression \eqref{eq:computePP}\;

     \uIf{$\hat{\mathds{P}}^{(t+1)} [T^{1: \ell } \leq Q-1] \geq \lambda$ and $|\omega_{\ell}| <  |\omega_{\ell + \mu}|$}{
    Proceed to Iteration 2 and consider $\omega_{\ell+1}$\;}
    \Else{Broadcast the selected $\kappa$ at this Iteration 1\;}
    \textbf{(Iteration 2)}
      Select new maximal clique $\kappa$   over expanded  window $\omega_{\ell+1}$\;
      Compute probability $\hat{\mathds{P}}^{(t+1)} [T^{1: \ell +1} \leq Q-1]$ using expression \eqref{eq:computePP}\;

     \uIf{$\hat{\mathds{P}}^{(t+1)} [T^{1: \ell +1} \leq Q-1] \geq \lambda$ and $|\omega_{\ell +1}| <  |\omega_{\ell + \mu}|$}{
     Proceed to Iteration 3 and consider  $\omega_{\ell+2}$\;
    }
    \uElseIf{ $\hat{\mathds{P}}^{(t+1)} [T^{1:\ell+1} \leq Q-1] \geq \lambda$ and $|\omega_{\ell+1}| =  |\omega_{\ell + \mu}|$ }{
     Broadcast the selected $\kappa$ at  this  Iteration 2\;
    }
    \Else{Broadcast the selected $\kappa$ at the previous Iteration 1\;}
\textbf{(Iteration 3)} Repeat  the steps of Iteration 2\;

\caption{Expanding Window IDNC (EW-IDNC) Algorithm} \label{alg:EW}
\end{algorithm}\DecMargin{1em}

At \emph{ Step 1 of  Iteration 2}, the EW-IDNC algorithm selects a  new maximal clique $\kappa$ over the expanded feasible window $\omega_{\ell+1}$.  At \emph{Step 2 of Iteration 2}, the  algorithm computes   the  probability  $\hat{\mathds{P}}^{(t+1)} [T^{1:\ell+1} \leq Q-1]$ using expression  \eqref{eq:computePP}. At \emph{Step 3 of Iteration 2}, the  algorithm  performs one of the following three steps.
\begin{itemize}
\item It   proceeds to Iteration 3 and considers window $\omega_{\ell+2}$, if $\hat{\mathds{P}}^{(t+1)} [T^{1:\ell+1} \leq Q-1] \geq \lambda$ and $|\omega_{\ell+1}| <  |\omega_{\ell + \mu}|$. This is  the case when the design criterion for the first $\ell+1$ video layers is satisfied and the  window can be further expanded.
 \item  It  broadcasts the selected $\kappa$ at this  Iteration 2, if $\hat{\mathds{P}}^{(t+1)} [T^{1:\ell+1} \leq Q-1] \geq \lambda$ and $|\omega_{\ell+1}| =  |\omega_{\ell + \mu}|$. This is  the case when the design criterion for the first $\ell+1$ video layers is satisfied but the  window is already the largest feasible window.\footnotemark \footnotetext{When the design criterion for the first $\ell +1$ video layers is satisfied, the design criterion for the first $\ell$ video layers is certainly satisfied since the number of missing packets of any receiver in the first $\ell$ video layers is smaller than or equal to that in the first $\ell +1$ video layers.}
 \item It  broadcasts the selected $\kappa$ at the previous Iteration 1, if $\hat{\mathds{P}}^{(t+1)} [T^{1:\ell+1} \leq Q-1] < \lambda$.  This is  the case when the design criterion for the first $\ell+1$ video layers is not satisfied.
\end{itemize}
At Iteration 3, the algorithm performs the steps of Iteration 2. This iterative process  is repeated until the algorithm reaches to the largest feasible window  $\omega_{\ell+\mu}$ or the design criterion for the video  layers in a given feasible window is not satisfied.  The proposed  EW-IDNC algorithm is summarized  in Algorithm \ref{alg:EW}.

%
%

\subsection{Non-overlapping Window Instantly Decodable Network Coding (NOW-IDNC) Algorithm}

Our proposed \emph{non-overlapping window instantly decodable network coding} (NOW-IDNC) algorithm  always selects  a maximal clique $\kappa$  over  the smallest feasible  window $\omega_{\ell}$ following   Section \ref{formulation} or Section  \ref{heuristic}.  This guarantees  the highest level of protection  to the most important  video layer, which has not  been decoded yet by all receivers. In fact, the video layers are broadcasted one after another  following their  decoding order in a non-overlapping manner. 


\section{Packet Selection Problem  over a Given Window} \label{formulation}
In this section, we address the problem of selecting  a maximal clique $\kappa$  over any given window $\omega_{\ell}$ that increases the possibility of decoding  those $\ell$ video layers   by the maximum number of receivers before the deadline. We first extract SFM $\mathbf F^{1:\ell}$ corresponding to window $\omega_{\ell}$ and construct IDNC graph $\mathcal G^{1:\ell}$ according to the extracted SFM $\mathbf F^{1:\ell}$. We then select a maximal clique $\kappa^*$  over  graph $\mathcal G^{1:\ell}$ in two stages. The packet selection problem  can be summarized  as follows.
\begin{itemize}
\item We  partition   IDNC graph $\mathcal G^{1:\ell}$ into critical graph $\mathcal G_c^{1:\ell}$ and non-critical graph $\mathcal G_b^{1:\ell}$. The critical graph $\mathcal G_c^{1:\ell}$  includes  the vertices generated from the missing packets in the first $\ell$ video layers at the critical receivers  in $\mathcal C^{1:\ell}$.   Similarly, the non-critical graph $\mathcal G_b^{1:\ell}$  includes the vertices generated from the missing packets in the first $\ell$ video layers at the  non-critical  receivers  in $\mathcal B^{1:\ell}$.
\item We prioritize  the critical receivers for the first $\ell$ video layers  over the non-critical receivers for the first $\ell$ video layers since  all the missing packets at the critical receivers cannot  be delivered  without targeting them in the current transmission (i.e., $W_i^{1:\ell} = Q, \forall R_i \in \mathcal C^{1:\ell}$).
\item If there is one or more critical receivers (i.e., $\mathcal{C}^{1:\ell} \neq \varnothing$), in the first stage,  we  select $\kappa_c^*$ to target  a subset of, or if possible, all critical receivers. We define   $\mathcal{X}_c$ as  the set of targeted critical receivers who have vertices in  $\kappa_c^*$.
\item If there is one or more non-critical receivers (i.e., $\mathcal{B}^{1:\ell} \neq \varnothing$), in the second stage, we select $\kappa_b^*$ to target a subset of, or if possible, all  non-critical receivers that do not violate the instant decodability constraint for the targeted critical receivers in $\kappa_c^*$. We define   $\mathcal{X}_b$ as  the set of targeted non-critical receivers who have vertices in  $\kappa_b^*$.
\end{itemize}

\subsection{Maximal Clique Selection Problem over Critical Graph}\label{sec:CriticalClique}

With  maximal  clique $\kappa_c^*$ selection,  each critical receiver  in $ \mathcal C^{1:\ell}(t)$  experiences one of the following two  events at time slot $t$:
\begin{itemize}
\item $R_i \in \mathcal{X}_{c}$,   the targeted critical receiver can still  receive  $W_i^{1:\ell}$ missing packets in the  exact $Q = W_i^{1:\ell}$ transmissions.
\item  $R_i \in \mathcal{C}^{1:\ell} \setminus \mathcal{X}_{c}$, the ignored critical receiver cannot  receive  $W_i^{1:\ell}$ missing packets in the remaining $Q-1$ transmissions and  becomes an affected receiver at time slot $t+1$.
\end{itemize}
Let $\mathcal{A}^{1:\ell}(t+1)$ be the  set of affected receivers for the first $\ell$ video layers at  time slot $t+1$ after $\kappa_{c}^*$ transmission at time slot $t$. The critical receivers  that are not targeted at time slot $t$ will become the new affected receivers, and the  critical receivers that  are targeted at time slot $t$ can also become the new affected receivers if they experience an erasure in this transmission. Consequently, we can express  the expected  increase in the number of affected receivers  from time slot $t$ to time slot $t+1$ after selecting $\kappa_c^*$  as:
\begin{align}
\mathds{E}[A^{1:\ell}(t+1) - A^{1:\ell}(t)] &=  (C^{1:\ell}(t) - |\mathcal{X}_{c}|) + \sum_{R_i \in  \mathcal{X}_{c}} \epsilon_i \nonumber \\
  &=  C^{1:\ell}(t) - \sum_{ R_i \in \mathcal{X}_{c}} 1 + \sum_{R_i \in  \mathcal{X}_{c}}\epsilon_i \nonumber \\
  &=  C^{1:\ell}(t) - \sum_{ R_i \in \mathcal{X}_{c}} (1 - \epsilon_i).
\end{align}
We now formulate  the problem of minimizing the expected increase in the   number of affected receivers for the first $\ell$ video layers from time slot $t$ to time slot $t+1$ as a critical   maximal clique selection problem  over critical  graph $\mathcal G_c^{1:\ell}$ such as:
\begin{align} \label{eqn:critical12}
\kappa_c^*(t) &= \arg \;\min_{\kappa_c \in \mathcal G_c^{1:\ell}} \left\{\mathds{E}[A^{1:\ell}(t+1) - A^{1:\ell}(t)] \right\} \nonumber \\
               &= \arg \;\min_{\kappa_c \in \mathcal G_c^{1:\ell}} \left \{ C^{1:\ell}(t) - \sum_{ R_i \in \mathcal{X}_{c}(\kappa_{c})} (1 - \epsilon_i) \right \}.
\end{align}
In other words, the problem of  minimizing the expected increase in the number of affected receivers is equivalent to finding all the maximal  cliques in the  critical IDNC graph, and selecting the maximal clique among them that results in the minimum expected increase in the   number of affected receivers.

\subsection{Maximal Clique Selection Problem over Non-critical Graph}
Once  maximal  clique $\kappa_c^*$ is selected  among the critical receivers in  $\mathcal{C}^{1:\ell}(t)$, there may exist vertices belonging to the non-critical receivers in non-critical graph $\mathcal G_b^{1:\ell}$ that can form  even a bigger maximal clique. In fact, if the selected new vertices are connected to all vertices in $\kappa_c^*$, the corresponding   non-critical receivers are targeted without affecting IDNC constraint for the targeted critical receivers in  $\kappa_c^*$. Therefore, we first extract non-critical subgraph  $\mathcal G_b^{1:\ell}(\kappa_c^*)$ of vertices in $\mathcal G_b^{1:\ell}$ that are adjacent to all the vertices in $\kappa_c^*$  and then select $\kappa_b^*$ over subgraph $\mathcal{G}_b^{1:\ell}(\kappa_c^*)$.

With these considerations, we aim to  maximize the upper bound on the probability that  individual  completion times of all non-critical receivers for the first $\ell$ video layers, starting from the successor time slot $t+1$, is less than or equal to the remaining $Q-1$ transmissions (represented by $\hat{\mathds P}^{(t+1)} [T_B^{1:\ell} \leq Q-1]$). We formulate this problem as  a non-critical maximal clique selection problem over   graph $\mathcal{G}_b^{1:\ell}(\kappa_c^*)$  such as:

\begin{align} \label{eqn:formulation3349}
\kappa_b^*(t)  = \arg \;\max_{\kappa_b \in \mathcal G_b^{1:\ell}(\kappa_c^*) } \left\{\hat{\mathds{P}}^{(t+1)} [T_B^{1:\ell}\leq Q-1]\right\}.
\end{align}

By  maximizing  probability $\hat{\mathds P}^{(t+1)} [T_B^{1:\ell} \leq Q-1]$ upon selecting  a maximal clique $\kappa_{b}$, the sender increases the probability of  transmitting all  packets in the first $\ell$ video layers to all non-critical receivers in $\mathcal B^{1:\ell}(t)$ before the deadline. Using   similar arguments   for non-critical receivers as in expression \eqref{eq:metric22},  we can define expression \eqref{eqn:formulation3349} as:
\begin{align} \label{eqn:formulation334}
\kappa_b^*(t)  =   \arg \;\max_{\kappa_b \in \mathcal G_b^{1:\ell}(\kappa_c^*)} \left\{ \prod_{R_i \in \mathcal{X}_{b}(\kappa_b)}  \mathds{P}^{(t)} [T_{W_i^{1:\ell}} \leq Q] \prod_{R_i \in \mathcal{B}^{1:\ell}(t) \setminus  \mathcal{X}_b(\kappa_b)}  \mathds{P}^{(t)} [T_{W_i^{1:\ell}} \leq Q-1]\right\}.
\end{align}
In other words, the problem of  maximizing  probability $\hat{\mathds{P}}^{(t+1)} [T_B^{1:\ell}\leq Q-1]$ for all non-critical receivers  is equivalent to finding all the maximal cliques in the non-critical subgraph $\mathcal{G}_b^{1:\ell}(\kappa_c^*)$, and selecting the maximal clique among them that results in the maximum  probability  $\hat{\mathds{P}}^{(t+1)} [T_B^{1:\ell}\leq Q-1]$.

\begin{remark}
\emph{The final served  maximal clique $\kappa^*$ over a given window $\omega_{\ell}$ is the union of two maximal cliques $\kappa_c^*$ and  $\kappa_b^*$  (i.e., $\kappa^* = \{\kappa_c^* \cup \kappa_b^* \}$).}
\end{remark}

It is well known that an $V$-vertex graph has  $O(3^{V/3})$ maximal cliques and  finding a maximal clique among them  is NP-hard  \cite{garey1979computers}. Therefore,  solving the formulated packet selection problem quickly leads to high computational complexity even for  systems with moderate numbers of receivers and packets ($V = O(MN)$). To  reduce the computational complexity, it is conventional to design an approximation algorithm.  However, the  problem is even hard to approximate since there is no $O(V^{1-\delta})$ approximation for the best maximal clique among $O(3^{V/3})$ maximal cliques  for any fixed $\delta > 0$ \cite{hastad1996clique}.

\ifCLASSOPTIONonecolumn
\vspace{-1mm}
\fi


\section{Heuristic Packet Selection  Algorithm  over a Given Window} \label{heuristic}
Due to the high computational complexity of the formulated packet selection  problem in Section \ref{formulation},   we now  design a low-complexity heuristic  algorithm following  the   problem formulations    in \eqref{eqn:critical12} and \eqref{eqn:formulation334}. This heuristic algorithm selects  maximal cliques $\kappa_c$ and  $\kappa_b$ based on a greedy  vertex search over IDNC graphs $\mathcal G_c^{1:\ell}$ and  $\mathcal G_b^{1:\ell}(\kappa_c)$, respectively.\footnotemark \footnotetext{Note that a similar greedy vertex search approach was studied in \cite{sorour2010completion,aboutorabenabling} due to its computational simplicity. However, the works in \cite{sorour2010completion,aboutorabenabling} solved different problems  and ignored the dependency between source packets and the hard deadline. These additional constraints considered in this paper lead us  to a different heuristic algorithm with its own features.}
\begin{itemize}
\item If there is one or more critical receivers (i.e., $\mathcal{C}^{1:\ell}(t) \neq \varnothing$), in the first stage, the algorithm  selects maximal clique $\kappa_c$ to reduce the  number of newly affected receivers for the first $\ell$ video layers after this transmission.
\item If there is one or more non-critical receivers (i.e., $\mathcal{B}^{1:\ell}(t) \neq \varnothing$), in the second stage, the algorithm selects maximal clique $\kappa_b$ to increase the probability $\hat{\mathds{P}}^{(t+1)} [T_B^{1:\ell}\leq Q-1]$ after this transmission.
\end{itemize}

\subsection{Greedy Maximal Clique Selection  over Critical Graph}\label{sec:CriticalHeuristic}
To select critical  maximal clique $\kappa_c$, the proposed algorithm starts by finding a lower bound on the potential new affected receivers, for the first $\ell$ video layers from time slot $t$ to time slot $t+1$,  that may result from selecting each  vertex from critical IDNC graph $\mathcal G_c^{1:\ell}$.  At Step 1, the algorithm selects  vertex $v_{ij}$ from  graph $\mathcal G_c^{1:\ell}$ and adds it to $\kappa_c$. Consequently,  the lower bound on the expected number of new affected  receivers for the first $\ell$ video layers  after this transmission that may result from selecting this vertex can be  expressed as:
\begin{equation}\label{eqn:LBstep1}
A^{1:\ell(1)}(t+1) -  A^{1:\ell}(t) =  C^{1:\ell}(t) - \sum_{R_m \in \{R_i \cup \mathcal M_{ij}^{\mathcal G_c^{1:\ell}} \} } (1-\epsilon_m).
\end{equation}
Here, $A^{1:\ell(1)}(t+1)$ represents the number of affected receivers for the first $\ell$ video layers at time slot $t+1$ after transmitting  $\kappa_c$ selected at Step 1  and $\mathcal M_{ij}^{\mathcal G_c^{1:\ell}}$ is the set of critical receivers that have at least one vertex adjacent to vertex $v_{ij}$ in $\mathcal G_c^{1:\ell}$. Once $A^{1:\ell(1)}(t+1) -   A^{1:\ell}(t)$   is calculated for all vertices in  $\mathcal G_c^{1:\ell}$, the algorithm chooses vertex $v_{ij}^*$ with the minimum lower bound on the expected number of  new affected receivers as:
\begin{equation} \label{eqn:vertexS1}
v_{ij}^* = \arg\min_{v_{ij} \in \mathcal G_c^{1:\ell}} \left\{A^{1:\ell(1)}(t+1) -   A^{1:\ell}(t) \right\}.
\end{equation}
After adding  vertex $v_{ij}^*$  to $\kappa_c$ (i.e., $\kappa_c = \{v_{ij}^* \}$), the algorithm extracts the subgraph $\mathcal G_c^{1:\ell}(\kappa_c)$ of vertices in $\mathcal G_c^{1:\ell}$ that are adjacent to all the  vertices in $\kappa_c$.
At Step 2, the algorithm selects another  vertex $v_{mn}$ from subgraph  $\mathcal G_c^{1:\ell}(\kappa_c)$ and adds it to $\kappa_c$. Consequently, the new lower bound on the expected number of new affected receivers can be expressed as:

\begin{align}  \label{distortionI}
A^{1:\ell(2)}(t+1) - A^{1:\ell}(t) &=   C^{1:\ell}(t)  - \left(\sum_{R_i \in \mathcal{X}_c(\kappa_c)} (1-\epsilon_i)   + \sum_{R_o \in \{R_m \cup \mathcal M_{mn}^{\mathcal G_c^{1:\ell}(\kappa_c)} \} } (1-\epsilon_o)\right) \nonumber \\
&=  \left( C^{1:\ell}(t) - \sum_{R_m \in \{R_i \cup \mathcal M_{ij}^{\mathcal G_c^{1:\ell}} \} } (1-\epsilon_m)\right) + \sum_{R_o \in \mathcal M_{ij}^{\mathcal G_c^{1:\ell}} \setminus  (R_m \cup  \mathcal M_{mn}^{\mathcal G_c^{1:\ell}(\kappa_c)})  } (1-\epsilon_o) \nonumber \\
&=  \left(A^{1:\ell(1)}(t+1) -  A^{1:\ell}(t)\right) + \sum_{R_o \in \{ \mathcal M_{ij}^{\mathcal G_c^{1:\ell}} \setminus  (R_m \cup  \mathcal M_{mn}^{\mathcal G_c^{1:\ell}(\kappa_c)}) \} } (1-\epsilon_o).
\end{align}

Since $(R_m \cup  \mathcal M_{mn}^{\mathcal G_c^{1:\ell}(\kappa_c)})$ is a subset of $\mathcal M_{ij}^{\mathcal G_c^{1:\ell}}$,  the last term in  \eqref{distortionI} is resulting from  the  stepwise increment on the lower bound on the expected number of newly affected receivers due to selecting  vertex $v_{mn}$. Similar to Step 1, once $A^{1:\ell(2)}(t+1) -   A^{1:\ell}(t)$ is calculated for all vertices in the  subgraph $\mathcal G_c^{1:\ell}(\kappa_c)$, the algorithm chooses vertex $v_{mn}^*$ with the minimum lower bound on the expected number of new affected receivers as:
\begin{equation} \label{eqn:vertex22}
v_{mn}^* = \arg\min_{v_{mn} \in \mathcal G_c^{1:\ell} (\kappa_c)} \{A^{1:\ell(2) }(t+1)- A^{1:\ell}(t) \}.
\end{equation}
After adding new vertex $v_{mn}^*$  to $\kappa_c$ (i.e., $\kappa_c = \{\kappa_c, v_{mn}^* \}$), the  algorithm repeats the vertex search process  until no further vertex in $\mathcal G_c^{1:\ell}$ is adjacent to all the vertices in $\kappa_c$.

\subsection{Greedy Maximal Clique Selection over Non-critical Graph}

To select non-critical  maximal clique $\kappa_b$, the proposed algorithm  extracts the non-critical  IDNC subgraph $\mathcal G_b^{1:\ell}(\kappa_c)$ of vertices in $\mathcal G_b^{1:\ell}$ that are adjacent to all the vertices in $\kappa_c$. This algorithm   starts  by finding the maximum  probability $\hat{\mathds{P}}^{(t+1)} [T_B^{1:\ell}\leq Q-1]$   that may result from selecting each vertex from subgraph  $\mathcal G_b^{1:\ell}(\kappa_c)$. At Step 1, the algorithm selects  vertex $v_{ij}$ from $\mathcal G_b^{1:\ell}(\kappa_c)$ and adds it to $\kappa_b$. Consequently,  the probability $\hat{\mathds{P}}^{(t+1),(1)} [T_B^{1:\ell}\leq Q-1]$  that may result from selecting this  vertex at Step 1  can be computed   as:
\begin{align} \label{eqn:CTstep1}
\hat{\mathds{P}}^{(t+1),(1)} [T_B^{1:\ell}\leq Q-1] &=   \prod_{R_m \in \{R_i \cup \mathcal M_{ij}^{\mathcal G_b^{1:\ell}(\kappa_c)}\}}  \mathds{P} [T_{W_m^{1:\ell}} \leq Q] \nonumber \\
 &  \prod_{R_m \in \{\mathcal B^{1:\ell}(t) \setminus (R_i \cup \mathcal M_{ij}^{\mathcal G_b^{1:\ell}(\kappa_c)}) \}}  \mathds{P} [T_{W_m^{1:\ell}} \leq Q-1].
\end{align}
Here, $\mathcal M_{ij}^{\mathcal G_b^{1:\ell}(\kappa_c)}$ is the set of non-critical receivers that have at least one vertex adjacent to vertex $v_{ij}$ in $\mathcal G_b^{1:\ell}(\kappa_c)$.
Once probability $\hat{\mathds{P}}^{(t+1),(1)} [T_B^{1:\ell}\leq Q-1] $ is calculated for all  vertices in   $\mathcal G_b^{1:\ell}(\kappa_c)$, the algorithm chooses vertex $v_{ij}^*$ with  the  maximum probability   as:
\begin{equation} \label{eqn:vertexS1}
v_{ij}^* = \arg\max_{v_{ij} \in \mathcal G_b^{1:\ell}(\kappa_c)} \{\hat{\mathds{P}}^{(t+1),(1)} [T_B^{1:\ell}\leq Q-1]\}.
\end{equation}
After adding   vertex $v_{ij}^*$  to $\kappa_b$ (i.e., $\kappa_b = \{v_{ij}^* \}$), the algorithm extracts the subgraph $\mathcal G_b^{1:\ell}(\kappa_c \cup \kappa_b)$ of vertices in $\mathcal G_b^{1:\ell}(\kappa_c)$ that are adjacent to all the  vertices in $(\kappa_c \cup \kappa_b)$. At Step 2,  the algorithm selects another vertex $v_{mn}$ from subgraph $\mathcal G_b^{1:\ell}(\kappa_c \cup \kappa_b)$ and adds it to $\kappa_b$.
Note that the new  set of potentially targeted non-critical receivers  after Step 2 is  $ \{R_i \cup R_m \cup \mathcal M_{mn}^{\mathcal G_b^{1:\ell}(\kappa_c \cup \kappa_b)}\}$, which is a  subset of $\{R_i  \cup \mathcal M_{ij}^{\mathcal G_b^{1:\ell}(\kappa_c)}\}$.  Consequently, the new  probability  $\hat{\mathds{P}}^{(t+1),(2)} [T_B^{1:\ell}\leq Q-1]$ due to the  stepwise reduction   in the number of targeted non-critical receivers  can be computed  as:
\begin{align} \label{eqn:CTstep}
\hat{\mathds{P}}^{(t+1),(2)} [T_B^{1:\ell}\leq Q-1] &=    \prod_{R_o  \in \{R_i \cup R_m \cup \mathcal M_{mn}^{\mathcal G_b^{1:\ell}(\kappa_c \cup \kappa_b)}\}}  \mathds{P} [T_{W_o^{1:\ell}} \leq Q] \nonumber \\
& \prod_{R_o \in \{\mathcal{B}^{1:\ell}(t)\setminus  (R_i \cup R_m \cup \mathcal M_{mn}^{\mathcal G_b^{1:\ell}(\kappa_c \cup \kappa_b)})\}}  \mathds{P} [T_{W_o^{1:\ell}} \leq Q-1].
\end{align}

\IncMargin{1em}
\begin{algorithm}[t]
Extract SFM $\mathbf F^{1:\ell}$ corresponding to a given window $\omega_{\ell}$\;
Construct $\mathcal G^{1:\ell}(\mathcal V,\mathcal E)$ according to the extracted SFM $\mathbf F^{1:\ell}$\;
Partition $\mathcal G^{1:\ell}$ into $\mathcal G_c^{1:\ell}$ and $\mathcal G_b^{1:\ell}$ according to the receivers in $\mathcal C^{1:\ell}$ and $\mathcal B^{1:\ell}$, repsectivley\;
Initialize $\kappa_c = \varnothing$ and $\kappa_b = \varnothing$\;

\While{$\mathcal G_c^{1:\ell} \neq \varnothing$}{
       Compute $A^{1:\ell}(t+1) - A^{1:\ell}(t), \forall v_{ij}\in \mathcal G_c^{1:\ell} (\kappa_c)$ using  \eqref{eqn:LBstep1} or \eqref{distortionI}\;
       Select $v_{ij}^*=\arg\min_{v_{ij}\in \mathcal G_c^{1:\ell}(\kappa_c)} \{A^{1:\ell}(t+1) - A^{1:\ell}(t)\}$\;
       Set $\kappa_c \leftarrow \kappa_c \cup v_{ij}^*$\;
       Update  subgraph $\mathcal G_c^{1:\ell}(\kappa_c)$ and  $\mathcal G_b^{1:\ell}(\kappa_c)$\;
}
\While{$\mathcal G_b^{1:\ell} \neq \varnothing $}{
      Compute  $\hat{\mathds{P}}^{(t+1)} [T_B^{1:\ell}\leq Q-1], \forall v_{ij}\in \mathcal G_b^{1:\ell}(\kappa_c \cup \kappa_b)$ using  \eqref{eqn:CTstep1} or \eqref{eqn:CTstep}\;
       Select $v_{ij}^*=\arg\max_{v_{ij}\in \mathcal G_b^{1:\ell}(\kappa_c \cup \kappa_b)}  \{ \hat{\mathds{P}}^{(t+1)} [T_B^{1:\ell}\leq Q-1] \}$\;
       Set $\kappa_b \leftarrow \kappa_b \cup v_{ij}^*$\;
       Update subgraph $\mathcal G_b^{1:\ell}(\kappa_c \cup \kappa_b)$\;
}
Set $\kappa \leftarrow \kappa_c\cup \kappa_b$.
\caption{Heuristic Packet Selection Algorithm over a Given Window} \label{alg:LGS}
\end{algorithm}\DecMargin{1em}

Similar to Step 1, once probability  $\hat{\mathds{P}}^{(t+1),(2)} [T_B^{1:\ell}\leq Q-1]$ is calculated for all vertices in the  subgraph $\mathcal G_b^{1:\ell}(\kappa_c \cup \kappa_b)$, the algorithm chooses vertex $v_{mn}^*$ with the maximum  probability   as:
\begin{equation} \label{eqn:CTstep2}
v_{mn}^* = \arg\max_{v_{mn} \in \mathcal G_b^{1:\ell}(\kappa_c \cup \kappa_b)} \{\hat{\mathds{P}}^{(t+1),(2)} [T_B^{1:\ell}\leq Q-1]\}.
\end{equation}
After adding  new vertex $v_{mn}^*$  to $\kappa_b$ (i.e., $\kappa_b = \{\kappa_b, v_{mn}^* \}$), the algorithm repeats the  vertex search process until no further vertex  in  $\mathcal G_b^{1:\ell}$ is adjacent to all the vertices in ($\kappa_c \cup \kappa_b$).

\begin{remark}
\emph{The  final maximal clique $\kappa$ is union of $\kappa_c$ and $\kappa_b$ (i.e., $\kappa =  \kappa_c \cup \kappa_b$). The proposed heuristic algorithm is summarized in Algorithm \ref{alg:LGS}. Note that  we  use this heuristic packet selection algorithm to select a maximal clique over a given window in EW-IDNC and NOW-IDNC algorithms  in Section \ref{results} (Simulation Results).}
\end{remark}

\begin{remark}
\emph{The  complexity of  the proposed  heuristic packet selection algorithm is $O(M^2N)$ since it requires weight computations for the $O(MN)$ vertices  in each step and  a maximal clique can have at most $M$ vertices. Using this heuristic algorithm,  the  complexity of  the   EW-IDNC algorithm is $O(M^2NL)$ since it can perform the heuristic algorithm at most $L$ times  over    $L$ windows. Moreover, using  this heuristic algorithm,  the  complexity of  the   NOW-IDNC algorithm is $O(M^2N)$ since it  performs the heuristic  algorithm once over the  smallest feasible window.}
\end{remark}

\ifCLASSOPTIONonecolumn
\vspace{-1mm}
\fi

\ifCLASSOPTIONonecolumn
\vspace{-1mm}
\fi
\section{Simulation Results over a Real Video Stream}\label{results}
In this section, we first discuss the scalable video test stream used in the simulation and then present the performances of   different algorithms for  that video stream.

\subsection{Scalable Video Test Stream}
\begin{figure}[t]
        \centering
        \includegraphics[width=8cm,height=5cm]{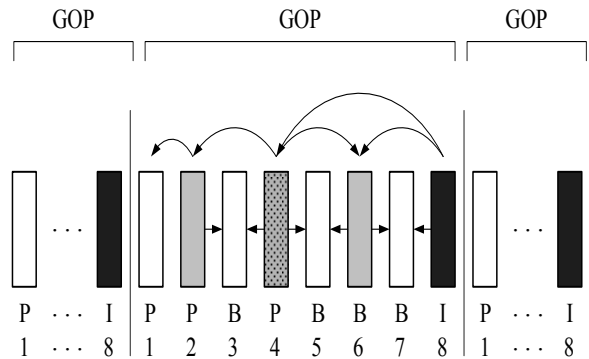}
        \caption{ A closed GOP with $4$ layers and $8$ frames  (a sequence of I, P and B frames).} \label{fig:GOP}
\end{figure}

We  now  describe the H.264/SVC video test stream used in this paper. We  consider a standard video stream,  Soccer \cite{test2014}. This stream is  in common intermediate format (CIF, i.e., $352 \times 288$) and has $300$ frames with $30$ frames per second. We encode the stream using the JSVM 9.19.14 version of H.264/SVC codec \cite{schwarz2007overview,software2011} while considering the GOP size of $8$ frames and temporal scalability of SVC.  As a result, there are $38$ GOPs for the test stream. Each GOP  consists of a sequence of I, P and B frames that are encoded into four video layers as shown in  Fig. \ref{fig:GOP}. The frames belonging to the same video layer are represented  by the identical  shade and   the more important  video layers are represented by the darker  shades.  In fact, the GOP in Fig. \ref{fig:GOP} is a closed GOP, in which the decoding of the frames inside the GOP is independent of frames outside the GOP \cite{esmaeilzadeh2014inter}. Based on the figure, we can see that a receiver can decode $1, 2, 4$ or $8$ frames upon receiving first 1, 2, 3 or 4 video layers, respectively. Therefore, nominal temporal resolution of 3.75, 7.5, 15 or 30 frames per second is experienced by a  viewer depending on the number of decoded  video layers.

To assign the information bits to  packets, we consider the maximum transmission unit (MTU) of $1500$ bytes as the size of a packet. We use $100$ bytes for  header information and  remaining $1400$ bytes for video data.  The average number of packets in the first, second, third and fourth video layers over  $38$ GOPs are $8.35, 3.11,  3.29$ and     $3.43$,  respectively.  For a GOP of interest, given that the number of frames per GOP is 8, the video frame rate is 30 frames per second, the transmission rate is $\alpha$ bit per second  and a packet length is $1500 \times 8$ bits, the allowable number of
transmissions  $\Theta$  for a GOP is fixed. We  can  conclude that $\Theta = \frac{8\alpha}{1500 \times 8 \times 30}$.

\subsection{Simulation Results }

We present the simulation results comparing the performance of our proposed EW-IDNC  and NOW-IDNC algorithms to the following algorithms.
\begin{itemize}
\item Expanding window RLNC (EW-RLNC) algorithm \cite{vukobratovic2012unequal,esmaeilzadeh2014inter} that uses  RLNC strategies to encode the packets  in different  windows while taking into account the decoding order of  video layers and the hard deadline. The encoding and decoding processes of EW-RLNC algorithm are described in Appendix \ref{app:EWRLNC}.
\item  Maximum clique (Max-Clique) algorithm \cite{le2013instantly} that  uses IDNC strategies to service a large number of receivers with any new packet in each transmission while ignoring the  decoding order of video layers and  the hard deadline.

\item  Interrelated priority encoding (IPE)   algorithm  \cite{wanginstantly} that uses IDNC strategies to  reduce the number of  transmissions   required  for  delivering the  base layer packets while  ignoring the hard deadline.
\end{itemize}

\begin{figure}[t]
        \centering
        \includegraphics[width=12cm,height=8.5cm]{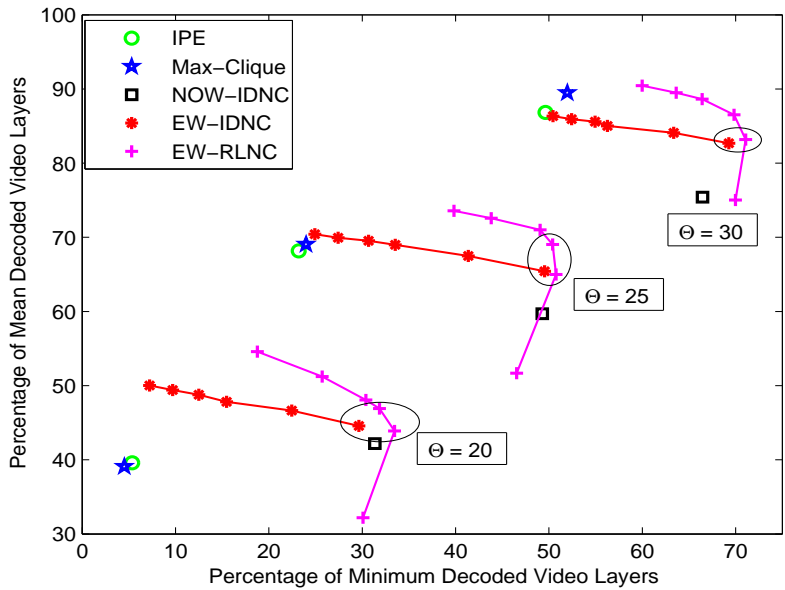}
        \caption{Percentage of mean  decoded video layers   versus percentage of minimum  decoded video layers  for different deadlines $\Theta$} \label{fig:deadlines}
\end{figure}

Figs.  \ref{fig:deadlines}  and \ref{fig:rx} show the percentage of  mean  decoded video layers  and the percentage of minimum  decoded video layers  performances of different algorithms for different deadlines $\Theta$ (for $M = 15, \epsilon = 0.2$) and  different numbers of receivers $M$ (for $\Theta = 25, \epsilon = 0.2$).\footnotemark \footnotetext{When  average erasure probability $\epsilon = 0.2$, the erasure probabilities of different receivers are in the range $[0.05,0.35]$. The simulation results are  the average based on  over 1000  runs.} We choose 6 values for threshold $\lambda$ from [0.2, 0.95] with step size of 0.15. This  results in 6 points on each trade-off curve of  EW-IDNC and EW-RLNC algorithms such as  $\lambda =0.2$  and  $\lambda = 0.95$ correspond to the top point and the bottom point, respectively. Moreover, we use ellipses to represent efficient operating points (i.e., thresholds $\lambda$) on the trade-off curves.  From both figures, we can draw the following observations:


\begin{itemize}
\item  As expected from EW-IDNC and EW-RLNC algorithms,  the minimum decoded video layers over all receivers increases  with the increase of threshold $\lambda$ at the expense of reducing the mean decoded video layers over all receivers. In general, given  a small threshold $\lambda$,  the design criterion is satisfied  for a large number of  video layers  in each transmission, which results in   a large coding  window and a low level of protection to the lower video layers. Consequently, several receivers may decode a large number of video layers, while other receivers may  decode only the first video layer before the deadline. To increase the minimum decoded video layers while respecting the  mean decoded video layers, an efficient  threshold $\lambda$ for the EW-IDNC algorithm is around $0.95$ and an efficient  threshold $\lambda$ for the EW-RLNC algorithm is around $0.65$.

\item  EW-RLNC algorithm performs poorly for  large thresholds (e.g., $\lambda = 0.95$ representing  the bottom point on the trade-off curve) due to transmitting a large number of coded packets from the smaller windows to obtain  high decoding probabilities of  the lower video layers at all receivers. Note that  EW-RLNC algorithm  explicitly determines the number of coded packets   from each window   at the beginning of the  $\Theta$ transmissions. In contrast,  our proposed  EW-IDNC algorithm  uses  feedbacks to determine an efficient coding window   in each transmission.

\begin{figure}[t]
        \centering
        \includegraphics[width=12cm,height=8.5cm]{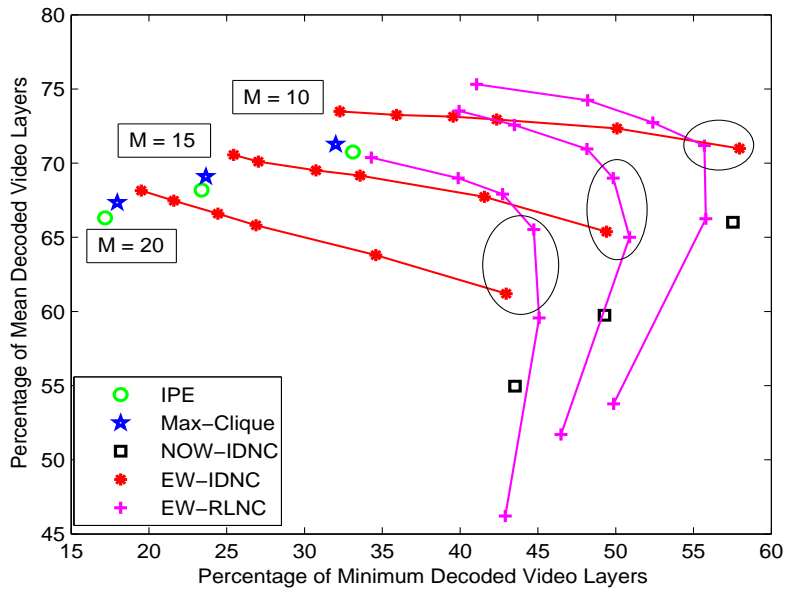}
        \caption{Percentage of mean  decoded video layers   versus percentage of minimum  decoded video
        layers for different number of receivers $M$} \label{fig:rx}
\end{figure}

\item  Our proposed EW-IDNC algorithm achieves   similar performances compared to the EW-RLNC algorithm  in terms of the minimum and the  mean decoded video layers. In fact, both algorithms  guarantee a high probability of completing the broadcast of a lower video layer (using  threshold $\lambda$)    before  expanding the  window over the  successor video layers.

 \item Our proposed NOW-IDNC algorithm  achieves a   similar performance compared to EW-IDNC and EW-RLNC algorithms in terms of the minimum decoded video layers.    However, the NOW-IDNC algorithm  performs poorly in terms of the mean decoded video layers due to always  selecting a packet combination over a single video layer.

\begin{figure}[t]
        \centering
        \includegraphics[width=12cm,height=8.5cm]{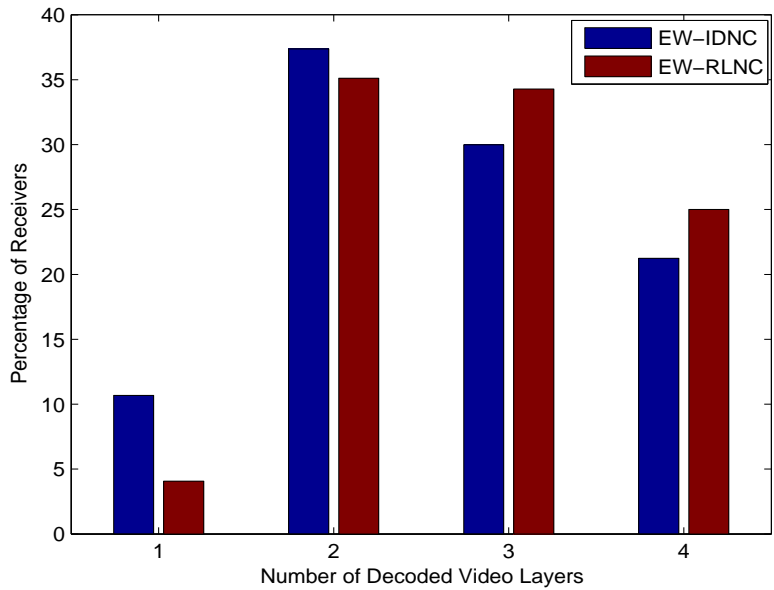}
        \caption{Histogram showing the percentage of receivers that successfully  decode one, two, three
        and four video layers before the deadline} \label{fig:hist}
\end{figure}

\item  As expected, Max-Clique and IPE algorithms perform poorly   compared to our proposed EW-IDNC and NOW-IDNC algorithms in terms of the minimum decoded video layers. Both Max-Clique and IPE algorithms make  coding decisions across all video layers and thus, do not address the hard deadline for the most important video layer. As a result,  several receivers may receive packets from the higher video layers, which  cannot be used for decoding those   video layers if a packet in a lower video layer is  missing  after the deadline.
\end{itemize}

Fig.  \ref{fig:hist} shows the histogram  obtained by  EW-IDNC algorithm (using $\lambda = 0.95$) and EW-RLNC algorithm (using $\lambda = 0.65$) for  $\Theta = 25,  M = 15, \epsilon = 0.2$. This histogram illustrates the  percentage of receivers that successfully decode one, two, three and four video layers before the deadline. From this histogram, we can see that most  of the receivers decode three or four  video layers out of four video  layers in a GOP. Moreover, the percentage of receivers that decode the first four video layers in  EW-RLNC algorithm   is slightly  higher  compared to that in EW-IDNC algorithm. This better performance of EW-RLNC algorithm  comes at the expense of higher packet overhead, higher encoding and decoding complexities as discussed in Section I.

\ifCLASSOPTIONonecolumn
\vspace{-1mm}
\fi
\section{Conclusion}\label{conclusion}
In this paper, we developed an efficient, yet computationally simple, IDNC framework for  real-time  scalable video broadcast  over wireless  networks. In particular, we  derived an upper bound on the probability that the individual completion times of all receivers  meet the deadline. Using this probability with  other  guidelines, we designed   EW-IDNC  and NOW-IDNC algorithms that provide a high level of protection to the most important  video layer before considering additional video layers in coding decisions.  We used a real scalable video stream in the simulation and showed that  our proposed  IDNC algorithms  improve the  received  video quality compared to the existing IDNC algorithms and achieve a similar performance compared to the EW-RLNC algorithm. Future research direction is to extend the proposed IDNC framework to  cooperative systems, where the receivers cooperate with  each other to recover their missing packets \cite{hernandez2014throughput}. In general, the short-range channels between the receivers are  better compared to the long-range channels between the base station  to the receivers, which can be beneficial for real-time  video streams with hard deadlines.

\begin{appendices}
 \ifCLASSOPTIONonecolumn
 \vspace{-1mm}
 \fi

\ifCLASSOPTIONonecolumn
\vspace{-1mm}
\fi
\section{Expanding Window Random Linear Network Coding} \label{app:EWRLNC}
We follow the work in \cite{esmaeilzadeh2014inter} and consider a deterministic approach, where the number of coded packets from each window is explicitly determined at the beginning of the  period of $\Theta$ transmissions. The sender broadcasts these coded packets in $\Theta$ transmissions without receiving  any feedback. Let us assume  that $\theta_{\ell}$ coded packets are generated (and thus transmitted) from the packets in the $\ell$-th window $\omega_{\ell}$. Then $\Sigma_{\ell=1}^L \theta_{\ell} = \Theta$ and  $\mathbf{z} = [\theta_1, \theta_2,...,\theta_L]$ is an EW-RLNC transmission policy. Given a fixed number of allowable transmissions $\Theta$,  all possible transmission policies can be defined as all combinations of the number of coded packets from each window. Now, we describe the process of selecting  a transmission  policy  as follows.

We use $\mathbf{n} = [n_1,n_2,...,n_L]$ to denote the number of packets from different layers in a GOP. For a given transmission policy $\mathbf{z}$, we denote the probability that  receiver $R_i$ with erasure probability $\epsilon_i$ can decode the packets of layer $\ell$ (and  all the packets of its lower layers) by $\mathds{P}_i^{\ell}(\mathbf{n},\mathbf{z})$. This probability can be computed  using  expression (1) in \cite{esmaeilzadeh2014inter}.
Now we extend this probability to $M$ receivers  and compute  the probability that  $M$ receivers  can decode the packets of layer $\ell$ (and  all the packets of its lower layers) as follows:
\begin{equation} \label{eqn:ewRLNC}
\mathds{P}^{\ell}(\mathbf{n},\mathbf{z})  = \prod_{R_i \in \mathcal{M}} \mathds{P}_i^{\ell}(\mathbf{n},\mathbf{z}).
\end{equation}
Given  transmission policy $\mathbf{z}$, the probability in \eqref{eqn:ewRLNC} is computed   for each of   $L$ video layers. Furthermore, we  consider all possible transmission policies and compute  probability $\mathds{P}^{\ell}(\mathbf{n},\mathbf{z}), \forall \ell \in [1,...,L]$, for each transmission policy. Finally, we select the transmission policy  $\mathbf{z}$ among all transmission policies that satisfies condition $\mathds{P}^{\ell}(\mathbf{n},\mathbf{z})  \geq  \lambda$ for the largest number of $\ell$ successive video layers (i.e., satisfies  condition for the largest $\ell$-th video layer and of course all its lower layers). Here,  condition $\mathds{P}^{\ell}(\mathbf{n},\mathbf{z})  \geq  \lambda$ is adopted following the same approach as in our proposed  EW-IDNC algorithm. The details of decoding a video layer based on the number of received packets from different windows can be found in \cite{esmaeilzadeh2014inter}.
\end{appendices}

\section*{Acknowledgment}
The authors would like to thank Mohammad Esmaeilzadeh
for his  comments in using a real  video test stream in this paper.

\bibliographystyle{IEEEtran}
\bibliography{ref}

\end{document}